\def\Im{{\text{Im}}\,}

\def\kF{k_{\text{F}}}

\def\vF{v_{\text{F}}}
\def\NF{N_{\text{F}}}

\def\kB{k_{\text{B}}}
\def\epsilonF{\epsilon_{\text F}}

\def\sgn{{\text{sgn\,}}}
\def\be{\begin{equation}}
\def\ee{\end{equation}}
\def\bea{\begin{eqnarray}}
\def\eea{\end{eqnarray}}
\def\bse{\begin{subequations}}
\def\ese{\end{subequations}}

\documentclass[prb,twocolumn,amsmath,amssymb,eqsecnum,preprintnumbers,longbibliography]{revtex4-1}
\usepackage{graphicx}
\usepackage{dcolumn}
\usepackage{bm}
\usepackage{bbm}
\usepackage{color}
\usepackage{soul}
\usepackage{enumitem}
\begin{document}
\preprint{arXiv:2409.08123}
\preprint{Phys. Rev. B {\bf 110}, 184203 (2024)}
\title{Diffusion, Long-Time Tails, and Localization in Classical and Quantum Lorentz Models: A Unifying Hydrodynamic Approach}
\author{T.R. Kirkpatrick$^1$ and D. Belitz$^{2,3}$}
\affiliation{ $^{1}$ Institute for Physical Science and Technology,
                    University of Maryland, College Park,
                    MD 20742, USA\\
                    $^{2}$ Department of Physics and Institute for Fundamental Science,
                    University of Oregon, Eugene, OR 97403, USA\\
                   $^{3}$ Materials Science Institute, University of Oregon, Eugene,
                    OR 97403, USA\\             
            }
\date{\today}

\begin{abstract}
Long-time tails, or algebraic decay of time-correlation functions, have long been known to exist both
in many-body systems and in models of non-interacting particles in the presence of quenched disorder
that are often referred to as Lorentz models. In classical Lorentz models, one manifestation is the long-time
tail of the velocity autocorrelation function. In quantum systems, the best known example is
what is known as weak-localization effects in disordered systems of non-interacting electrons. A wide
variety of techniques have been used to study these phenomena, including kinetic theory and mode-coupling
theories for classical systems, and many-body theory and field-theoretic techniques for quantum mechanical
ones. This paper provides a unifying, and very simple, approach to these effects. We show that simple
modifications of the diffusion equation due to either a random diffusion coefficient, or a random
scattering potential, accounts for both the decay exponents and the prefactors of the leading 
long-time tails in the velocity autocorrelation functions of both classical and quantum Lorentz models.

\end{abstract}


\maketitle

\section{Introduction}
\label{sec:I}

Kinetic theory at the level of the Boltzmann equation predicts that equilibrium time correlation functions
generically decay exponentially in the limit of long times. This seemingly plausible prediction turned
out to not be true. As was found first by means of computer simulations,\cite{Alder_Wainwright_1970} and
shortly thereafter understood theoretically,\cite{Dorfman_Cohen_1970, Ernst_Hauge_van_Leeuwen_1970}
the generic long-time behavior is described by power laws known as long-time tails (LTTs). The LTTs are 
generic in the sense that they appear everywhere in the phase diagram, in contrast to, e.g., the critical
dynamics observed at isolated critical points. Their physical
origin are so-called `memory effects' that arise from repeated collisions of the same particles. As a 
result,  not all collisions are statistically independent, whereas Boltzmann theory assumes they are.
This leads to correlations that are long-ranged in time.
Technically, LTTs arise from wave-number integrals over soft, or massless, correlation functions,
which makes them intrinsically dependent on the spatial dimensionality $d$, with a lower dimensionality
resulting in a stronger LTT for a given system. In the system studied originally,\cite{Alder_Wainwright_1970}
namely, a hard-sphere fluid, the velocity autocorrelation function decays as $t^{-d/2}$ for long times.
The same is true for a realistic classical fluid.\cite{Ernst_Hauge_van_Leeuwen_1976a, Ernst_Hauge_van_Leeuwen_1976b}
In a simpler model known as a classical Lorentz gas, which consists of a single classical particle
moving in a random array of hard-sphere scatterers, the LTT of the velocity autocorrelation function
is weaker, $t^{-(d+2)/2}$, and has the opposite sign.\cite{Ernst_Weyland_1971, LTT_sign_footnote} 
More recently, LTTs have attracted renewed interest in a wide variety of contexts, including
supersymmetric field theories and quantum gravity,\cite{Kovtun_Yaffe_2003, Caron-Huot_Saremi_2010} 
nuclear physics,\cite{Shukla_2021} quantum chaos,\cite{Abbasi_Tabatabaei_2020, Abbasi_2022}
and the role of hydrodynamic fluctuations in thermalization.\cite{Lin_Delacretaz_Hartnoll_2019}

A related phenomenon with a very similar physical origin is the fact that transport coefficients do
not allow for an analytic density expansion in analogy to the virial expansion for thermodynamic
quantities.\cite{Weinstock_1965, Dorfman_Cohen_1965, Dorfman_Cohen_1967} Rather, at a 
certain order in an expansion in powers of the density, which depends on the spatial dimensionality, 
the analog of the virial coefficient diverges logarithmically.\cite{Sengers_1965, Kawasaki_Oppenheim_1965} 
In a classical Lorentz model the expansion parameter is the dimensionless scatterer density $n_i a^d$, with 
$n_i$ the scatterer density and $a$ the scattering length (which in a hard-sphere model is just the scatterer 
radius). In a quantum Lorentz model, often also referred to as the Anderson model, which describes 
non-interacting electrons in an environment of random scatterers,\cite{Anderson_1958} the leading expansion parameter is 
$n_i a^{d-1} \lambda$, with $\lambda$ the de Broglie wavelength of the scattered 
particle.\cite{Kirkpatrick_Dorfman_1983, Wysokinski_et_al_1994, Wysokinski_et_al_1995} In
either case the logarithm appears at linear order in the expansion parameter for $d=2$, and at
quadratic order for $d=3$. The same parameters, $n_i a^d$ and $n_i a^{d-1} \lambda$, 
appear as the prefactors of the LTTs in classical and quantum Lorentz models, as we will
emphasize in Sec.~\ref{sec:III} and \ref{sec:IV}. Alternatively, the quantum expansion
parameter can be written as $1/\kF\ell$ or $1/\epsilonF\tau$, with $\kF$ and $\epsilonF$ the Fermi 
wave number and energy, and $\ell$ and $\tau$ the mean-free path and mean-free
time, respectively, see Eq.~(\ref{eq:1.3}).

The existence of LTTs has far-reaching consequences. Transport coefficients are given as
time integrals over appropriate time correlation functions,\cite{Dorfman_vanBeijeren_Kirkpatrick_2021}
and the algebraic long-time decay of the latter leads to a non-analytic frequency dependence of the 
former.\cite{Pomeau_1972} If the LTT is strong enough the result may be that a transport coefficient 
does not exist since the time correlation function is not integrable. For instance, the $t^{-d/2}$ LTT in a
classical fluid implies that the viscosity and the heat conductivity do not exist in $d\leq 2$, and the usual 
hydrodynamic description of the fluid breaks down.\cite{Forster_Nelson_Stephen_1977} In general, 
the physical meaning of these strong LTTs is not clear {\em a priori}. In a $2$-$d$ quantum Lorentz 
model the LTT is as strong as in a classical fluid, but with the opposite sign. That is, the perturbative 
correction to the diffusion coefficient is negative and diverges logarithmically with decreasing frequency
or temperature. This result is often called `weak localization'.\cite{Bergmann_1984} It implies a breakdown 
of perturbation theory and its ultimate physical meaning is a vanishing diffusion coefficient
for any nonzero impurity density.\cite{Abrahams_et_al_1979, Wegner_1979} In $3$-$d$ Lorentz models,
both quantum and classical, the static diffusion coefficient for small disorder is suppressed by a 
finite amount compared to its Boltzmann value. This decrease is accompanied by an increase 
as a function of the frequency that is nonanalytic, corresponding to a LTT. With increasing
impurity density the diffusion coefficient further decreases and eventually vanishes at a
critical impurity density. In quantum systems, this diffusion blocking is know as Anderson
localization.\cite{Anderson_1958, Lee_Ramakrishnan_1985, Evers_Mirlin_2008}
There is a vast amount of literature on these problems, and a large number of different
techniques have been used to study them, including kinetic theory,\cite{Ernst_Weyland_1971} 
scaling arguments,\cite{Abrahams_et_al_1979} mode-coupling theories,\cite{Ernst_Hauge_van_Leeuwen_1971,
Ernst_Hauge_van_Leeuwen_1976a, Ernst_Hauge_van_Leeuwen_1976b, Goetze_1979, Goetze_Leutheusser_Yip_1981, 
Goetze_1981} many-body diagrams,\cite{Gorkov_Larkin_Khmelnitskii_1979, Vollhardt_Woelfle_1980}
and nonlinear sigma models.\cite{Wegner_1979, Schaefer_Wegner_1980, Evers_Mirlin_2008}

In this paper we provide a very simple unifying approach based on basic hydrodynamic
arguments that is capable of describing LTTs in both classical and quantum Lorentz models.
We will show that fluctuations of the impurity density lead to effects that are physically
distinct from fluctuations of the scattering potential, with the leading LTTs in classical systems
caused by the former, while the latter dominate in quantum systems. The leading LTTs in $d=2,3$
spatial dimensions are as follows. In a classical Lorentz gas, spatial fluctuations of the scatterer
density lead to fluctuations of the diffusion coefficient that result in a LTT for the velocity time-correlation
function
\bse
\label{eqs:1.1}
\be
\lim_{t\to\infty} \langle{\bm v}(t)\cdot{\bm v}(t=0)\rangle \propto - (n_i a^d)^{d-1} (t/\tau)^{-(d+2)/2}\ .
\label{eq:1.1a}
\ee
A fluctuating scattering potential leads to a LTT with the same exponent, but a subleading
prefactor,
\be
\lim_{t\to\infty} \langle{\bm v}(t)\cdot{\bm v}(t=0)\rangle \propto - (n_i a^d)^{d+1} (t/\tau)^{-(d+2)/2}\ .
\label{eq:1.1a}
\ee
\ese
Quantum mechanically, the importance of these two mechanisms is reversed: A fluctuating
scattering potential leads to the weak-localization result
\bse
\label{eqs:1.2}
\be
\lim_{t\to\infty} \langle{\bm v}(t)\cdot{\bm v}(t=0)\rangle \propto - (n_i a^{d-1}\lambda)^{d-1} (t/\tau)^{-d/2}\ ,
\label{eq:1.2a}
\ee
whereas a fluctuating diffusion coefficient leads to the same exponent but a subleading prefactor,
\be
\lim_{t\to\infty} \langle{\bm v}(t)\cdot{\bm v}(t=0)\rangle \propto - (n_i a^d)^{d-1} (n_i a^{d-1}\lambda)^2 (t/\tau)^{-d/2}\ .
\label{eq:1.2b}
\ee
\ese
We note again that in the weak-localization literature the coupling constant  $n_i a^{d-1}\lambda$ 
is usually written as $1/\kF\ell$ or $1/\epsilonF\tau$ (see Eq.~(\ref{eq:3.9}) for the relation between
$a$ and $\tau$):
\be
n_i a^{d-1}\lambda = 2\pi/\kF\ell = \pi\hbar/\epsilonF\tau\ .
\label{eq:1.3}
\ee
We will use these ways to write the coupling constant interchangeably to some extent. 
For comparison we also list the LTT for the shear viscosity in a classical fluid. Here the relevant 
time-correlation function is the autocorrelation of the shear stress $\sigma_{\perp}$,
\be
\lim_{t\to\infty} \langle \sigma_{\perp}(t)\,\sigma_{\perp}(t=0)\rangle \propto + (n a^d)^{d-1} (t/\tau)^{-d/2}\ ,
\label{eq:1.4}
\ee
with $n$ the fluid number density. The LTT exponent is the same as for weak localization,
Eq.~(\ref{eq:1.2a}), but the sign is different and the prefactor is not necessarily small.

This paper is organized as follows. In Sec.~\ref{sec:II} we discuss modifications of the
diffusion equation due to spatial fluctuations of (1) the diffusion coefficient, and (2) the
scattering potential. In Sec.~\ref{sec:III} we show that fluctuations of the diffusion
coefficient due to the random impurity density are responsible for the leading LTT
in a classical Lorentz model. The fluctuating potential leads to a LTT that has been
derived before using various methods, but is not the leading LTT in either classical
or quantum systems. In Sec.~\ref{sec:IV} we show how to use an inherently quantum 
mechanical symmetry to obtain the leading quantum LTT that leads to weak localization,
and in Sec.~\ref{sec:V} we conclude with a discussion. Appendix~\ref{app:A} recalls 
the relation between low-frequency non-analyticities and LTTs, Appendix~\ref{app:B}
demonstrates the consistency between different ways to obtain the LTTs,
Appendix~\ref{app:C} explains the relation between our treatment of the 
fluctuating-potential problem and mode-coupling theories, and Appendix~\ref{app:D}
discusses quantum effects on the LTT that results from a fluctuating diffusion
coefficient.

\section{Fluctuating Diffusion Equation}
\label{sec:II}

The basic observable we will discuss is the space and time-dependent particle number
density $n({\bm x},t)$. We will consider non-interacting particles that move in $d$ spatial 
dimensions ($d=2,3$) in a random array of static, or quenched, elastic scatterers, or impurities, that
provide a random potential $U({\bm x})$ and whose average density is $n_i$. Classically, 
the simplest system of this kind is a Lorentz gas, i.e., a single particle with fixed velocity 
$v_0$ that moves among hard-sphere scatterers. Quantum mechanically, we consider 
non-interacting electrons with the Fermi velocity given by $\vF$. Generically we will denote
the particle velocity by ${\bm v}$, whose modulus $v$ is either $v_0$ or $\vF$ depending 
on whether we consider the classical or the quantum case. 

The conserved nature of $n$ is expressed by a continuity equation
\be
\partial_t\, n({\bm x},t) = -{\bm\nabla}\cdot{\bm j}({\bm x},t)\ ,
\label{eq:2.1}
\ee
with ${\bm j}$ the number current density. As long as ${\bm j}$ is not specified, Eq.~(\ref{eq:2.1}) is an exact statement.
In order to obtain a closed equation for $n$, the standard phenomenological assumption is that ${\bm j}$ is proportional
to minus the gradient of the quantity conjugate to $n$, i.e, the chemical potential $\mu$:
\be
{\bm j} = -\sigma {\bm\nabla}\mu = -\sigma \left(\frac{\partial\mu}{\partial n}\right)_{V,T} {\bm\nabla} n\ .
\label{eq:2.2}
\ee
Here $\sigma$ is an Onsager coefficient, $V$ is the system volume, $T$ is the temperature, and we assume
that there is no temperature gradient. Combining Eqs.~(\ref{eq:2.2}) and (\ref{eq:2.1}) we obtain
\be
\partial_t\,n = {\bm\nabla}\cdot \left(D\,{\bm\nabla}n\right)
\label{eq:2.3}
\ee
with $D = \sigma/(\partial n/\partial\mu)_{T,V}$ a diffusion coefficient.

At the level of the Boltzmann equation, the diffusion coefficient is a constant $D_0$
and we obtain the standard diffusion equation
\bse
\label{eqs:2.4}
\be
\partial_t\,\delta n({\bm x},t) = D_0 {\bm\nabla}^2 \delta n({\bm x},t)\ ,
\label{eq:2.4a}
\ee
with $\delta n$ the deviation of the density from its average value $n_0$. After a spatial Fourier transform this reads
\be
(\partial_t + D_0 k^2)\, \delta n({\bm k},t) = 0\ .
\label{eq:2.4b}
\ee
\ese
The LTTs to be discussed in this paper all result from disorder-induced modifications of Eq.~(\ref{eq:2.4b}) that lead 
to nonanalytic frequency dependent renormalizations of the diffusion coefficient. Since the diffusion coefficient is the 
Laplace transform of the velocity auto-correlation function,\cite{Dorfman_vanBeijeren_Kirkpatrick_2021}
\be
D(z) = \frac{1}{d} \int_0^{\infty} dt\,e^{izt}\,\langle {\bm v}(t)\cdot{\bm v}(t=0)\rangle\quad (\text{for}\ \text{Im}(z)>0)\ ,
\label{eq:2.5}
\ee
such nonanalyticities imply a non-exponential decay of the correlation function for $t\to\infty\ $ \cite{Pomeau_1972}
via the relevant Tauberian theorems.\cite{Lighthill_1958, Feller_1970}
 
These modifications of the diffusion equation
have two distinct physical origins: Quenched disorder leads to (1) a random position 
dependence of the diffusion coefficient, and (2) a random scattering potential. As we will see, 
the effects resulting from the two mechanisms are qualitatively different. The former is 
responsible for the leading LTT in a classical Lorentz gas, while the latter yields the leading 
LTT in quantum systems.

\subsection{Fluctuating diffusion coefficient}
\label{subsec:II.A}

Due to the randomly distributed impurities, the diffusion coefficient $D$ in Eq.~(\ref{eq:2.3})
is position dependent, and a random variable. If we replace it by its spatial average $D_0$ we 
obtain Eq.~(\ref{eq:2.4a}). More generally, we write 
$D({\bm x}) = D_0 + \delta D({\bm x})$. Let $n_0$ be the average density, and
write $n({\bm x},t) = n_0 + \delta n({\bm x},t)$. The fluctuating diffusion equation then takes the form
\bse
\label{eqs:2.6}
\be
\partial_t\,\delta n({\bm x},t) - D_0 {\bm\nabla}^2 \delta n({\bm x},t) = {\bm\nabla} \delta D({\bm x}) \cdot{{\bm\nabla}}\delta n({\bm x},t)\ ,
\label{eq:2.6a}
\ee
or, after a spatial Fourier transform,
\be
\left(\partial_t + D_0 k^2\right)\delta n({\bm k},t) = \frac{-1}{V}\sum_{\bm q} ({\bm k}\cdot{\bm q})\,\delta D({\bm k}-{\bm q})\,\delta n({\bm q},t)\ .
\label{eq:2.6b}
\ee
A temporal Laplace transform with complex frequency $z$ yields
\bea
\delta n({\bm k},z)&=& {\frak D}({\bm k},z)\,\delta n({\bm k},t=0)
\nonumber\\
&& - {\frak D}({\bm k},z)\,\frac{1}{V} \sum_{\bm q} ({\bm k}\cdot{\bm q})\,\delta D({\bm k}-{\bm q})\,\delta n({\bm q},z)\ .
\nonumber\\
\label{eq:2.6c}
\eea
\ese
Here ${\frak D}$ is a bare diffusion propagator which for $\text{Im}\, z > 0$ has the form
\be
{\frak D}({\bm k},z) = \frac{1}{-iz + D_0 k^2}\ .
\label{eq:2.7}
\ee
More generally,
\bse
\label{eqs:2.8}
\be
{\frak D}({\bm k},z) = \frac{1}{-iz + D_0 k^2 s(z)}\ ,
\label{eq:2.8a}
\ee
where
\be s(z) = \sgn\Im z\ .
\label{eq:2.8b}
\ee
For later reference we note that wave-number integrals over the diffusion propagator have the form
\be
\frac{1}{V}\sum_{\bm k} {\frak D}({\bm k},z) \equiv F(z) = s(z)\,f_d(-iz s(z))\ ,
\label{eq:2.8c}
\ee
\ese
where the function $f_d$ depends on the spatial dimension. $F(z)$ has the properties
\bse
\label{eqs:2.9}
\bea
F(-z) &=& -F(z)\ ,
\label{eq:2.9a}\\
F(z^*) &=& -F(z)^*\ ,
\label{eq:2.9b}
\eea
\ese
which reflect the fact that the underlying velocity autocorrelation function is real and an even function of time.

We still need to specify the statistical properties of the random variable $\delta D$. We assume
that it is Gaussian distributed with zero mean,
\bse
\label{eqs:2.10}
\be
\{\delta D({\bm x})\}_{\text{dis}} = 0
\label{eq:2.10a}
\ee 
and second moment
\be
\{\delta D({\bm x})\,\delta D({\bm y})\}_{\text{dis}} = \Delta({\bm x}-{\bm y})\ ,
\label{eq:2.10b}
\ee
or, in Fourier space,
\be
\{\delta D({\bm k})\,\delta D({\bm p})\}_{\text{dis}} = V \delta_{{\bm k},-{\bm p}}\,\Delta({\bm k})\ .
\label{eq:2.10c}
\ee
\ese
Here $\{\ldots\}_{\text{dis}}$ is the disorder average, and rotational invariance on average implies
that $\Delta({\bm k})$ depends on the modulus $k$ of ${\bm k}$ only. 

A simple realization of a fluctuating diffusion coefficient relies on the fluctuations of the impurity
density only. Simple kinetic arguments yield $D_0 = (v^2/d)\tau$ in terms of the mean-free time 
$\tau \propto 1/n_i$ that is inversely proportional to the scatterer density. In general, the latter is
a fluctuating quantity $n_i({\bm x})$. If we replace it by its spatial average $n_i$ we obtain the 
Boltzmann or Drude value $\tau_0$ for the mean-free time, and the corresponding value 
$D_0 = (v^2/d)\tau_0$ for the diffusion coefficient that appears in the standard diffusion equation 
(\ref{eq:2.4a}). More generally, we write $n_i({\bm x}) = n_i + \delta n_i({\bm x})$, which leads to a 
fluctuating diffusion coefficient
$D({\bm x}) = D_0 + \delta D({\bm x})$, with
\be
\delta D({\bm x}) = - (D_0/n_i)\delta n_i({\bm x})\ .
\label{eq:2.11}
\ee
Assuming a Gaussian distribution for $\delta n_i$ with second moment
\bse
\label{eqs:2.12}
\bea
\left\{\delta n_i({\bm x})\,\delta n_i({\bm y})\right\}_{\text{dis}} &=& n_i\,\delta({\bm x}-{\bm y})\ ,
\label{eq:2.12a}\\
\left\{\delta n_i({\bm k})\,\delta n_i({\bm p})\right\}_{\text{dis}} &=& n_i\,V\,\delta_{{\bm k},-{\bm p}}\ ,
\label{eq:2.12b}
\eea
\ese
we obtain
\be
\Delta({\bm k}) = D_0^2/n_i
\label{eq:2.13}
\ee
for the second moment in Eq.~(\ref{eq:2.10c}). In this simple model, $\Delta({\bm k})$ is
independent of the wave number. That is, the fluctuations of the diffusion coefficient are
delta-correlated in real space. More generally, one expects the correlations to extend
over a microscopic distance on the order of the inverse Fermi wave number. This will
cause $\Delta$ to fall off for $k\agt \kF$; this will be important for discussing a particular 
quantum effect in Appendix~\ref{app:D}.

We stress that the above considerations rely on basic statistics only, are very general, 
and make no reference to the nature of the scattering process. As such, their consequences
must be the same in both classical and quantum mechanical systems.

\subsection{Fluctuating scattering potential}
\label{subsec:II.B}

Now consider a random static scattering potential $U({\bm x})$ and an associated random force
${\bm F}({\bm x}) = -{\bm\nabla}U({\bm x})$. The resulting acceleration of a particle
of mass $m$ is ${\bm a} = {\bm F}/m$, which leads to a drift velocity ${\bm u} = {\bm a}\tau$, and
hence to a contribution to the current density given by
\be
\delta{\bm j} = {\bm u}\,\delta n = \frac{-\tau}{m}\,({\bm\nabla}U)\,\delta n\ .
\label{eq:2.14}
\ee
The continuity equation (\ref{eq:2.3}) for the density thus gets modified to
\be
\partial_t \delta n({\bm x},t) = {\bm\nabla}\cdot D({\bm x}) {\bm\nabla} \delta n({\bm x},t) + \frac{\tau}{m}\,{\bm\nabla}\cdot \left[({\bm\nabla} U({\bm x}))\,\delta n({\bm x},t)\right]\ .
\label{eq:2.15}
\ee
If we ignore the fluctuations of the diffusion coefficient this becomes
\bse
\label{eqs:2.16}
\be
\partial_t \delta n({\bm x},t) = D_0 {\bm\nabla}^2 \delta n({\bm x},t) + \frac{\tau}{m}\,{\bm\nabla}\cdot \left[({\bm\nabla} U({\bm x}))\,\delta n({\bm x},t)\right]\ .
\label{eq:2.16a}
\ee
Note that the additional contribution to the current density in Eqs.~(\ref{eq:2.14}) and (\ref{eq:2.16a}) is
proportional to the gradient of the random potential. This is in contrast to Eq.~(\ref{eq:2.6a}), where
the corresponding contribution is proportional to the gradient of $\delta n$. After a spatial Fourier
transform Eq.~(\ref{eq:2.16a}) becomes
\be
(\partial_t + D_0 k^2)\delta n({\bm k},t) = - \frac{\tau}{m}\,\frac{1}{V}\sum_{\bm q} ({\bm k}\cdot{\bm q})\,U({\bm q})\,\delta n({\bm k}-{\bm q},t)\ ,
\label{eq:2.16b}
\ee
and a temporal Laplace transform yields
\bea
\delta n({\bm k},z) &=& {\frak D}({\bm k},z)\,\delta n({\bm k},t=0)
\nonumber\\ && - {\frak D}({\bm k},z)\,\frac{\tau}{m}\,\frac{1}{V}\sum_{\bm q} ({\bm k}\cdot{\bm q})\,U({\bm q})\,\delta n({\bm k}-{\bm q},z)\ .
\nonumber\\
\label{eq:2.16c}
\eea
\ese

We assume that the random potential $U$ is Gaussian distributed with zero mean,
\bse
\label{eqs:2.17}
\be
\{U({\bm q})\}_{\text{dis}} = 0\ ,
\label{eq:2.17a}
\ee
and second moment
\be
\{U({\bm q})\,U({\bm q}')\}_{\text{dis}} = V \delta_{{\bm q},-{\bm q}'}\,{\cal U}^2({\bm q})\ .
\label{eq:2.17b}
\ee
Classically, we model the correlation ${\cal U}^2$ by
\be
{\cal U}^2({\bm q}) = (n_i\, U^2/q_0^{2d}) \Theta(q_0 - k)\ .
\label{eq:2.17c}
\ee
Here $U^2$ is a measure of the potential strength, and $1/q_0$ is the range of the potential correlations. 
Quantum mechanically we assume pure s-wave scattering, in which case we have\cite{Vollhardt_Woelfle_1980}
\be
{\cal U}^2({\bm q}) \equiv {\cal U}_0 = \hbar/2\pi\NF\tau\ ,
\label{eq:2.17d}
\ee
\ese
with $\NF$ the density of states at the Fermi level. Note that this potential correlation corresponds 
to a quantum mechanical scattering process that vanishes in the classical limit.\cite{QM_footnote}
In a kinetic-theory treatment it is equivalent to keeping only those
contributions to the collision operator that are of lowest order in the scattering length.\cite{Kirkpatrick_Dorfman_1983}
Indeed, we will see that it leads to a coupling constant for the LTT that is given by
$n_i a^{d-1} \lambda$, with $\lambda$ the de Broglie wavelength of the scattered particle. In addition,
there are contributions of higher order in the scattering length, leading to a coupling constant $n_i a^d$, which we neglect.

In this generalization of the diffusion equation quantum mechanics enters via the correlations
of the random potential. Equation~(\ref{eq:2.17d}) is appropriate for a quantum Lorentz model.\cite{Kirkpatrick_Dorfman_1983}
In the classical limit, $\hbar\to 0$, the right-hand side of Eq.~(\ref{eq:2.17d}) vanishes as $\hbar^{d+1}$
and Eq.~(\ref{eq:2.17c}) provides a simple model for the potential correlations.

\section{Classical and Semi-Classical Long-Time Tails}
\label{sec:III}

In this section we calculate the density-density time-correlation function 
\bea
S_{nn}({\bm k},t) = \frac{1}{V}\,\langle \delta n({\bm k},t)\,\delta n(-{\bm k},t=0)\rangle\ .
\label{eq:3.1}\\
\nonumber
\eea
to obtain the LTT of the diffusion coefficient due to the quenched disorder. Here 
$\langle \ldots\rangle = \{\langle\ldots\rangle_{\text{ens}}\}_{\text{dis}}$ comprises both an 
ensemble average and the disorder average (see Appendix~\ref{app:B.1} for a discussion of
the ensemble average). Multiplying Eq.~(\ref{eq:2.4b}) with $\delta n(-{\bm k},t=0)$ and
averaging yields
\bse
\label{eqs:3.2}
\be
\left(\partial_t + D_0\, k^2\right) S_{nn}({\bm k},t) = 0\ ,
\label{eq:3.2a}
\ee
or, after a Laplace transform,
\be
\left(-iz + D_0 k^2\right) S_{nn}({\bm k},z) = S_{nn}({\bm k},t=0)\ .
\label{eq:3.2b}
\ee
The equal-time correlation function on the right-hand side is a well-behaved function of ${\bm k}$.
To leading order in $k^2$ and $z$ we therefore can write
\be
S_{nn}({\bm k},z) = {\frak D}({\bm k},z)\,S_{nn}^{\,0}\ ,
\label{eq:3.2c}
\ee
\ese
where $S_{nn}^{\,0} = S_{nn}({\bm k}=0,t=0)$. In what follows we will routinely use this replacement.

Since the diffusive nature of $S_{nn}$ is a consequence of the conservation law, Eq.~(\ref{eq:2.1}),
this functional form must be preserved if one starts with Eq.~(\ref{eq:2.6b}) or (\ref{eq:2.16b}) instead,
and the effects of the quenched disorder must be restricted to renormalizing the diffusion coefficient. 

\medskip
\subsection{Long-time tail due to a fluctuating diffusion coefficient}
\label{subsec:III.A}

\subsubsection{Correction to the diffusion coefficient}
\label{subsubsec:III.A.1}

Multiplying Eq.~(\ref{eq:2.6b}) with $\delta n(-{\bm k},t=0)$ and averaging 
we obtain
\begin{widetext}
\bse
\label{eqs:3.3}
\be
\left(\partial_t + D_0\, k^2\right) S_{nn}({\bm k},t) = \frac{-1}{V^2} \sum_{\bm q} ({\bm k}\cdot{\bm q})\,\langle\delta D({\bm k}-{\bm q}) \delta n({\bm q},t)\,\delta n(-{\bm k},t=0)\rangle
\label{eq:3.3a}
\ee
or, after a temporal Laplace transform,
\be
\left(-iz + D_0 k^2\right) S_{nn}({\bm k},z) = S_{nn}^{\,0} - \frac{1}{V^2} \sum_{\bm q} ({\bm k}\cdot{\bm q})\,\langle\delta D({\bm k}-{\bm q}) \delta n({\bm q},z)\,\delta n(-{\bm k},t=0)\rangle\ .
\label{eq:3.3b}
\ee
\ese
Inserting Eq.~(\ref{eq:2.6c}) on the right-hand side we obtain
\bea
\left(-iz + D_0 k^2\right) S_{nn}({\bm k},z) &=& S_{nn}^{\,0}
      - \frac{1}{V^2} \sum_{\bm q} ({\bm k}\cdot{\bm q})\,{\frak D}({\bm q},z) \langle \delta D({\bm k}-{\bm q})\,\delta n({\bm q},t=0)\,\delta n(-{\bm k},t=0)\rangle
\nonumber\\
&& \hskip 20pt + \frac{1}{V^3} \sum_{{\bm q},{\bm q}'} ({\bm k}\cdot{\bm q})({\bm q}\cdot{\bm q}')\,{\frak D}({\bm q},z) \langle \delta D({\bm k}-{\bm q})\,\delta D({\bm q}-{\bm q}')\, \delta n({\bm q}',z)\,\delta n(-{\bm k},t=0)\rangle\ .\qquad
\label{eq:3.4}
\eea
\end{widetext}

Given Eq.~(\ref{eq:2.6a}), Eq.~(\ref{eq:3.4}) is a formally exact expression for the time correlation function $S_{nn}$. For small disorder,
to leading order in a disorder expansion, we can factorize the correlation functions on the right-hand side: 
$\langle \delta D\,\delta n\,\delta n\rangle \approx \{\delta D\}_{\text{dis}}\langle\delta n\,\delta n\rangle = 0$
and $\langle \delta D\,\delta D\,\delta n\,\delta n\rangle \approx \{\delta D\,\delta D\}_{\text{dis}} \langle\delta n\,\delta n\rangle$.
In this leading approximation we find, using Eq.~(\ref{eq:2.10c}),
\bea
\left(-iz + D_0 k^2\right) S_{nn}({\bm k},z) &\approx& S_{nn}^{\,0} \hskip 50pt
\nonumber\\
&& \hskip -100pt + \frac{1}{V}\sum_{\bm q}({\bm k}\cdot{\bm q})^2 \Delta({\bm k}-{\bm q})\,{\frak D}({\bm q},z)\,S_{nn}({\bm k},z)\ .\qquad
\label{eq:3.5}
\eea
We we see that the density-density correlation function in the long-wavelength limit retains its diffusive form, as it must, 
and the disorder renormalizes the diffusion coefficient by producing a 
frequency dependent correction $\Delta D(z)$ to the bare diffusion coefficient $D_0$:
\be
\Delta D(z) = \frac{-1}{dV} \sum_{\bm q} \Delta({\bm q})\,{\bm q}^2\,{\frak D}({\bm q},z)\ .
\label{eq:3.6}
\ee
The correction to the diffusion coefficient involves a wave-number integral over a diffusion pole, which leads
to a nonanalytic frequency dependence, and hence to a LTT. With Eq.~(\ref{eq:2.13}) for $\Delta({\bm q})$ 
we finally have
\be
\Delta D(z) = \frac{-1}{d}\,\frac{D_0^2}{n_i}\,\frac{1}{V}\sum_{\bm q} q^2\,{\frak D}({\bm q},z)\ .
\label{eq:3.7}
\ee
The same result can be obtained by considering the decay of an externally created density
perturbation, see Appendix~\ref{app:B.4.a}.

\subsubsection{Long-time tail of the velocity autocorrelation function}
\label{subsubsec:III.A.2}

Equation~(\ref{eq:3.7}) is identical with the result obtained from kinetic theory by Ernst and Weyland.\cite{Ernst_Weyland_1971, typo_footnote}
The same result was reproduced in Ref.~\onlinecite{Goetze_Leutheusser_Yip_1981} by means of a mode-coupling theory. The
integral over the diffusion pole leads to a non-analyticity of the frequency-dependent diffusion coefficient at $z = 0$.
The leading nonanalytic contribution is
\bse
\label{eqs:3.8}
\bea
\frac{\Delta D(z)}{D_0} &=& -(n_i a^d)^{d-1}\,A_d \hskip 120pt
\nonumber\\
&& \hskip -20pt \times (-iz\tau)^{d/2} \times \begin{cases} 1 & \text{for}\quad d=3\\
                                                                                                                        \ln(-i z \tau) & \text{for}\quad d=2
                                                                                                 \end{cases}     \ ,
\label{eq:3.8a}
\eea
where $A_d$ is a dimensionless coefficient given by
\be
A_d =  d^{(d-2)/2}/4\pi\ ,
\label{eq:3.8b}
\ee
\ese
and $a$ is the scattering length defined by 
\be
\tau = 1/n_i v a^{d-1}\ .
\label{eq:3.9}
\ee
Since the diffusion coefficient is the Laplace transform of the velocity auto-correlation function, 
see Eq.~(\ref{eq:2.5}), 
this nonanalyticity implies a non-exponential decay of the correlation function for $t\to\infty$. The relevant
Tauberian theorem (see  Appendix~\ref{app:A}, which is based on Refs.~\onlinecite{Lighthill_1958, Feller_1970}) yields,
for $d=2,3$,\cite{Ernst_Weyland_1971}
\bse
\label{eqs:3.10}
\bea
\lim_{t\to\infty}\langle {\bm v}(t)\cdot{\bm v}(t=0)\rangle &=& \frac{D_0^2}{n_i}\frac{-2d\pi}{(4\pi D_0 t)^{(d+2)/2}}\qquad\qquad
\label{eq:3.10a}\\
&& \hskip -100pt =  - v^2\,(n_i a^d)^{d-1}\,B_d\, (t/\tau)^{-(d+2)/2}\ . 
\label{eq:3.10b}
\eea
where
\be
B_d = (d/\pi)^{d/2}/2^{d+1}\ .
\label{eq:3.10c}
\ee
\ese
The result as given in Eq.~(\ref{eq:3.10b}) shows that the coupling constant for the
LTT is the dimensionless scatterer density $n_i a^d$, see Eq.~(\ref{eq:3.8a}).

\begin{widetext}
\subsection{Long-Time Tail due to a Fluctuating Potential}
\label{subsec:III.B}

\subsubsection{Correction to the diffusion coefficient}
\label{subsubsec:III.B.1}

Now we repeat the procedure from Sec.~\ref{subsubsec:III.A.1}, but for the case of a random
scattering potential. Equation~(\ref{eq:2.16b}) implies 
\bse
\label{eqs:3.11}
\bea
(\partial_t + D_0 k^2) S_{nn}({\bm k},t) 
         &=& - \frac{\tau}{m}\,\frac{1}{V^2}\sum_{\bm q} ({\bm k}\cdot{\bm q})\langle U({\bm q})\,\delta n({\bm k}-{\bm q},t)\,\delta n(-{\bm k},t=0)\rangle
\nonumber\\
&&\hskip -0pt = - \frac{\tau}{m}\,\frac{1}{V^2}\sum_{\bm q} ({\bm k}\cdot{\bm q})\langle U({\bm q})\,\delta n({\bm k}-{\bm q},t=0)\,\delta n(-{\bm k},-t)\rangle
\nonumber\\
&&\hskip -0pt = - \frac{\tau}{m}\,\frac{1}{V^2}\sum_{\bm q} ({\bm k}\cdot{\bm q})\langle U({\bm q})\,\delta n({\bm k}-{\bm q},t=0)\,\delta n(-{\bm k},t)\rangle\ .
\label{eq:3.11a}
\eea
In going from the first line to the second line in Eq.~(\ref{eq:3.11a}) we have used time translational
invariance, and in going to the third line we have assumed time reversal symmetry. The latter
assumption is not necessary; the same result is obtained in the absence of time reversal, but
at the expense of a more elaborate calculation, see Appendix~\ref{app:B.2}.

After a Laplace transform, Eq.~(\ref{eq:3.11a}) becomes
\be
(-iz + D_0 k^2) S_{nn}({\bm k},z) = S_{nn}^{\,0} - \frac{\tau}{m}\,\frac{1}{V^2} \sum_{\bm q} ({\bm k}\cdot{\bm q})\langle (U({\bm q})\,\delta n({\bm k}-{\bm q},t=0)\,\delta n(-{\bm k},z)\rangle\ .
\label{eq:3.11b}
\ee
\ese
Now we use Eq.~(\ref{eq:2.16c}) in the second term on the right-hand side. This yields
\bea
(-iz + D_0 k^2) S_{nn}({\bm k},z) &=& S_{nn}^{\,0} - {\frak D}({\bm k},z)\,\frac{\tau}{m}\,\frac{1}{V^2} \sum_{\bm q} ({\bm k}\cdot{\bm q})
        \langle U({\bm q})\,\delta n({\bm k}-{\bm q},t=0)\,\delta n(-{\bm k},t=0)\rangle 
\nonumber\\
&& \hskip 19pt - {\frak D}({\bm k},z)\,\frac{\tau^2}{m^2}\,\frac{1}{V^3} \sum_{{\bm q},{\bm q}'} ({\bm k}\cdot{\bm q})\,({\bm k}\cdot{\bm q}')
       \langle U({\bm q})\,U({\bm q}')\,\delta n({\bm k}-{\bm q},t=0)\,\delta n(-{\bm k}-{\bm q}',z)\rangle\ .  \qquad\quad
       \label{eq:3.12}
\eea
\end{widetext}
Inspecting the terms on the right-hand side of Eq.~(\ref{eq:3.12}) we see that the second term does not 
contain an integral over a diffusion pole. It therefore does not lead to a LTT, and we drop it. In the third
term, to leading order in the random potential, we can factorize the average and use Eq.~(\ref{eq:3.2b})
for the resulting density-density correlation function. We then obtain
\bea
\left(-iz + D_0 k^2\right) S_{nn}({\bm k},z) &=& S_{nn}^{\,0} \hskip 50pt
\nonumber\\
&& \hskip -100pt + \frac{\tau^2}{m^2}\,\frac{1}{V}\sum_{\bm q}({\bm k}\cdot{\bm q})^2\, {\cal U}^2({\bm q})\,{\frak D}({\bm q},z)\,S_{nn}({\bm k},z)\ .\qquad
\label{eq:3.13}
\eea
For the correction to the diffusion coefficient we now have
\be
\Delta D(z) = \frac{-\tau^2}{dm^2}\,\frac{1}{V} \sum_{\bm q} {\cal U}^2({\bm q})\,{\bm q}^2\,{\frak D}({\bm q},z)
\label{eq:3.14}
\ee
With the s-wave scattering expression for the potential, Eq.~(\ref{eq:2.17d}), this yields
\be
\Delta D(z) = \frac{-1}{4\pi}\,\frac{D_0^2}{n_e}\,\frac{\hbar}{\epsilonF\tau}\,\frac{1}{V}\sum_{\bm q} q^2\,{\frak D}({\bm q},z)\ 
\label{eq:3.15}
\ee
with $n_e$ the electron density. This result was first derived in Ref.~\onlinecite{Goetze_1979} by a different method, see Appendix~\ref{app:C}.

\subsubsection{Long-time tail of the velocity autocorrelation function}
\label{subsubsec:III.B.2}

Comparing Eqs.~(\ref{eq:3.15}) and (\ref{eq:3.7}) we see that the LTT is the same as in the case of a
fluctuating diffusion coefficient, just with a different prefactor. We find
\bse
\label{eqs:3.16}
\bea
\frac{\Delta D(z)}{D_0} &=& -(n_i a^{d-1}\lambda)^{d+1}\,A_d
\nonumber\\
&& \hskip -30pt \times (-iz\tau)^{d/2} \times \begin{cases} 1 & \text{for}\quad d=3\\
                                                                                                                        \ln(-i z \tau) & \text{for}\quad d=2
                                                                                                 \end{cases}\ ,  \qquad\qquad  
\label{eq:3.16a}
\eea
with
\be
A_d = d^{(d+2)/2}/2^{d+3}\pi^4\ .
\label{eq:3.16b}
\ee
\ese
For the LTT of the velocity autocorrelation function this yields (see Appendix~\ref{app:A})
\bse
\label{eqs:3.17}
\bea
\lim_{t\to\infty}\langle {\bm v}(t)\cdot{\bm v}(t=0)\rangle &=& - v^2\,(n_i a^{d-1}\lambda)^{d+1}\,B_d
\nonumber\\
&& \hskip 0pt \times\,(t/\tau)^{-(d+2)/2}\ .
\label{eq:3.17a}
\eea
with
\be
B_d = d^{2+d/2}/2^{2d+2}\,\pi^{3+d/2}\ .
\label{eq:3.17b}
\ee
\ese
For a classical scattering potential, Eq.~(\ref{eq:2.17c}), the prefactor $(n_i a^{d-1}\lambda)^{d+1}$ gets replaced by
an expression proportional to $(n_i a^d)^{d+1}(U/mv^2)^2$. Note that the s-wave scattering cross section is the only 
way quantum mechanics enters the result shown in Eqs.~(\ref{eqs:3.16}), (\ref{eqs:3.17}). It affects only the prefactor,
otherwise this contribution to the LTT is the same as the classical contribution in Sec.~\ref{subsec:III.A}, and we will
refer to it as the semi-classical LTT.
The dimensionless coupling constant, $n_i a^{d-1} \lambda$, is the same as in the density expansion for the quantum
Lorentz model,\cite{Kirkpatrick_Dorfman_1983, Wysokinski_et_al_1995} and it is larger than the
classical coupling constant $n_i a^d$ provided $\lambda > a$. However, in contrast to the case of
the density expansion, here the quantum mechanical prefactor appears with two additional powers of the coupling constant
compared to the classical one, as can be seen by comparing Eqs.~(\ref{eqs:3.16}, \ref{eqs:3.17}) with
Eqs.~(\ref{eqs:3.8}, \ref{eq:3.10b}); a term of order $(n_i a^{d-1}\lambda)^{d-1}$, which one might
naively expect, is absent. In the classical limit, $\lambda \ll a$, one needs to keep
the next-to-leading term in an expansion in powers of the scattering length
to obtain a contribution that is of the same form as Eqs.~(\ref{eqs:3.17}) but with a prefactor of 
$(n_i a^d)^{d+1}$.\cite{Kirkpatrick_Dorfman_1983}  This term represents a higher order correction to Eq.~(\ref{eq:3.10b}). 

The LTT given in Eqs.~(\ref{eqs:3.17}) is not the leading quantum contribution; interference
effects lead to a LTT that is a strong as in a classical fluid, i.e., $t^{-d/2}$, albeit with a
different sign. We discuss this in Section~\ref{sec:IV}.

\section{Quantum Long-Time Tails: Weak Localization}
\label{sec:IV}

The quantum LTT displayed in Eqs.~(\ref{eqs:3.16}) or (\ref{eqs:3.17}) is not the leading LTT
in a quantum Lorentz model. This is because the only quantum mechanical (QM) ingredient is the
s-wave scattering cross section, whereas the diffusion propagator ${\frak D}$ is classical
in nature: from Eq.~(\ref{eq:3.2c}) we see that it is the normalized van Hove function $S_{nn}$.
The first step for finding the QM generalizations of the LTTs discussed in
Sec.~\ref{sec:III} is therefore to find the appropriate QM density-density correlation. 

\subsection{Quantum mechanical density correlation function}
\label{subsec:IV.A}

To this end we recall the fluctuation-dissipation theorem, which relates the van Hove
function to the commutator correlation function, or susceptibility, $\chi_{nn}$,\cite{Forster_1975}
which is defined as
\bse
\label{eqs:4.1}
\be
\chi_{nn}({\bm k},t) = \frac{i}{\hbar\,V} \left\langle\left[\delta n({\bm k},t),\delta n(-{\bm k},t=0)\right]_-\right\rangle\ .
\label{eq:4.1a}
\ee
Here $\delta n$ is the operator-valued particle-number density fluctuation, $[\ ,\ ]_-$ denotes
a commutator, and the average contains a QM expectation value in addition to the 
statistical mechanics and disorder averages. With temporal Fourier and Laplace transforms
as defined in Appendix~\ref{app:A.1} we have
\be
\chi_{nn}({\bm k},z) = \int \frac{d\omega}{\pi}\,\frac{\chi_{nn}''({\bm k},\omega)}{\omega - z}\ ,
\label{eq:4.1b}
\ee
where
\be
\chi_{nn}''({\bm k},\omega) = \frac{1}{2i}\,\chi_{nn}({\bm k},\omega)
\label{eq:4.1c}
\ee
\ese
is the spectral density.\cite{notation_footnote} In the classical limit, the fluctuation-dissipation theorem states
\bse
\label{eqs:4.2}
\be
S_{nn}({\bm k},\omega) = \frac{2T}{\omega}\,\chi_{nn}''({\bm k},\omega)\ .
\label{eq:4.2a}
\ee
The equal-time van Hove function thus is
\be
S_{nn}^0 = T\,\chi_{nn}^0\ ,
\label{eq:4.2b}
\ee
with \be
\chi_{nn}^0 = \lim_{{\bm k}\to 0} \int \frac{d\omega}{\pi}\,\frac{\chi_{nn}''({\bm k},\omega)}{\omega}
\label{eq:4.2c}
\ee
\ese
the static susceptibility. For noninteracting electrons at $T=0$, $\chi_{nn}^0 = \NF$. 
For the Laplace transform this yields
\bse
\label{eqs:4.3}
\be
S_{nn}({\bm k},z)/S_{nn}^0 = \frac{-i}{\chi_{nn}^0}\,\Phi_{nn}({\bm k},z)\ ,
\label{eq:4.3a}
\ee
with
\be
\Phi_{nn}({\bm k},z) = \frac{1}{z}\,\left[\chi_{nn}({\bm k},z) - \chi_{nn}^0\right]
\label{eq:4.3b}
\ee
\ese
the density-density Kubo function. Comparing with Eq.~(\ref{eq:3.2c}) we see that the
desired QM generalization of the correction to the diffusion coefficient, Eq.~(\ref{eq:3.15}), is\cite{uniqueness_footnote}
\be
\Delta D(z) = \frac{i}{4\pi}\,\frac{D_0^2}{n_e \chi_{nn}^0}\,\frac{\hbar}{\epsilonF\tau}\,\frac{1}{V} \sum_{\bm q} q^2\,\Phi_{nn}({\bm q},z)\ .
\label{eq:4.4}
\ee

\subsection{The semi-classical long-time tail revisited}
\label{subsec:IV.B}

\begin{figure}[t]
\includegraphics[width=8.5cm]{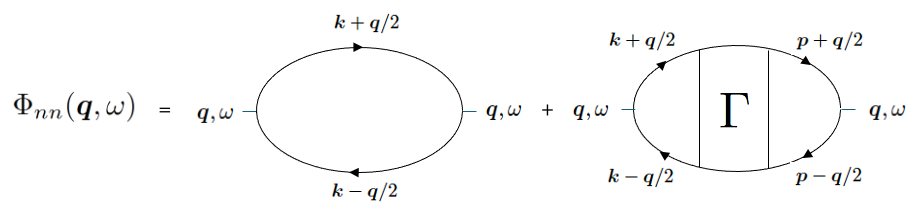}
\caption{Diagrammatic representation of the density-density Kubo function. Directed lines represent Green functions.}
\label{fig:1}
\end{figure}

\begin{figure}[t]
\includegraphics[width=8.5cm]{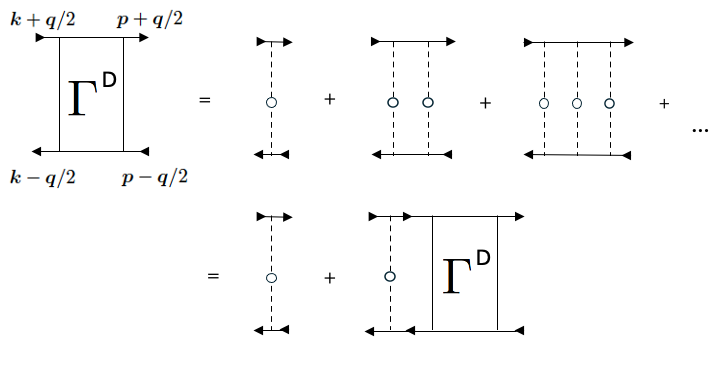}
\caption{Diagrammatic realization of the diffusion propagator. Dashed lines represent impurity potentials, 
              and circles represent factors of the impurity
              density; two dashed lines linked by a circle represent the factor ${\cal U}_0$.}
\label{fig:2}
\end{figure}

The result expressed in Eq.~(\ref{eq:3.15}) is semi-classical in the sense that the prefactor is QM in nature,
but the diffusive propagator relies on particle-number conservation only. This contribution to $\Delta D$
therefore must also exist in a fully QM treatment. To see that it does, we evaluate Eq.~(\ref{eq:4.4}) for
non-interacting electrons at $T=0$. Then $\chi_{nn}^0 = \NF$, and the leading contribution to the Kubo 
function in the limit of small frequencies and wave numbers can be written\cite{Vollhardt_Woelfle_1980}
\begin{widetext}
\bea
\Phi_{nn}({\bm q},\omega + i0) &=& \frac{1}{V^2} \sum_{{\bm k},{\bm p}}  \Phi_{{\bm k},{\bm p}}({\bm q},\omega)
\nonumber\\
               &=&  \frac{-1}{2\pi i} \sum_{{\bm k},{\bm p}} G_{{\bm k}+{\bm q}/2}(\omega + i0)\,G_{{\bm p}-{\bm q}/2}(-i0)
                        \left[V\,\delta_{{\bm k},{\bm p}}
                            + \Gamma_{{\bm k},{\bm p}}({\bm q},\omega)\,G_{{\bm p}+{\bm q}/2}(\omega + i0)\,G_{{\bm k}-{\bm q}/2}(-i0)\right]\ .\qquad
\label{eq:4.5}
\eea
\end{widetext}
Diagrammatically, this expression is shown in Fig.~\ref{fig:1}.
Here
\be
G_{\bm k}(z) = \frac{1}{z - \xi_{\bm k} \pm i/2\tau}\qquad (\text{$\pm$ for Im\,$z \genfrac{}{}{0pt}{2}{>}{<} 0$)}
\label{eq:4.6}
\ee
is the electron Green function with $\xi_{\bm k} = \epsilon_{\bm k} - \epsilonF$ in terms of the single-electron
energy $\epsilon_{\bm k}$, and $\omega \pm i0$
stands for $\lim_{\epsilon\to 0} \omega \pm i\epsilon$. $\Gamma$ is the impurity vertex function.
Consider the contribution $\Gamma^{\text{D}}$ to $\Gamma$ given by the sum of ladder diagrams shown in Fig.~\ref{fig:2}.
\begin{figure}[t]
\includegraphics[width=8.5cm]{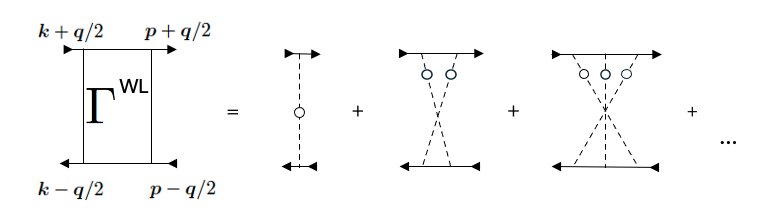}
\caption{Crossed-ladder approximation for the density vertex.}
\label{fig:3}
\end{figure}
In the limit of small $q$ and $\omega$ this is analytically\cite{Vollhardt_Woelfle_1980}
\be
\Gamma^{\text{D}}_{{\bm k},{\bm p}}({\bm q},\omega) = \frac{{\cal U}_0/\tau}{-i\omega + D_0 q^2}\ .
\label{eq:4.7}
\ee
The corresponding contribution to the Kubo function is
\be
\Phi^{\text{D}}({\bm q},\omega+ i0) = i\,\NF\,{\cal D}({\bm q},\omega+i0)\ .
\label{eq:4.8}
\ee
Using this result in Eq.~(\ref{eq:4.4}), and analytically continuing to complex frequencies $z$,
we recover Eq.~(\ref{eq:3.15}).

\subsection{The leading quantum long-time tail}
\label{subsec:IV.C}

Now consider the contribution $\Gamma^{\text{WL}}$ (with `WL' standing for `weak localization') to the 
vertex function $\Gamma$ shown in Fig.~\ref{fig:3}, which consists of all maximally crossed impurity lines. 
The calculation yields\cite{Gorkov_Larkin_Khmelnitskii_1979, Vollhardt_Woelfle_1980}
\be
\Gamma^{\text{WL}}_{{\bm k},{\bm p}}({\bm q},\omega) = \frac{{\cal U}_0/\tau}{-i\omega + D_0 ({\bm k}+{\bm p})^2}\ .
\label{eq:4.9}
\ee
This is valid in the limit $\omega\to 0$, ${\bm k}+{\bm p}\to 0$. It is important to note that the diffusive nature
of $\Gamma^{\text{WL}}$, in contrast to $\Gamma^{\text{D}}$ in Eq.~(\ref{eq:4.7}), is not protected by
particle number conservation. The corresponding contribution to the Kubo function is, for $\omega\to 0$,
\bea
\Phi^{\text{WL}}({\bm q},\omega) &=& \frac{-1}{2\pi i}\,\frac{{\cal U}_0}{\tau}\,\frac{1}{V}\sum_{\bm p} {\cal D}({\bm p},z)\,
\nonumber\\
&& \times \frac{1}{V}\sum_{\bm k}
     \vert G_{{\bm k}+{\bm q}}(i0)\vert^2\, \vert G_{{\bm k}}(i0)\vert^2\ .\qquad
\label{eq:4.10}
\eea
Using the $T=0$ results
\bse
\label{eqs:4.11}
\bea
\frac{1}{V} \sum_{\bm k} \vert G_{\bm k}(i0)\vert^2 &=& 1/{\cal U}_0\ ,
\label{eq:4.11a}\\
\frac{1}{V}\sum_{\bm k} k^2\, \vert G_{\bm k}(i0)\vert^2 &=& \kF^2/{\cal U}_0\ ,
\label{eq:4.11b}
\eea
\ese
in Eq.~(\ref{eq:4.10}) we find for the correction to the diffusion coefficient from Eq.~(\ref{eq:4.4})
\be
\Delta D^{\text{WL}}(z) = \frac{-1}{\pi\hbar\NF}\,\frac{1}{V}\sum_{\bm q} {\frak D}({\bm q},z)\ .
\label{eq:4.12}
\ee
Note that the above arguments yield the exact prefactor for the weak-localization LTT, in addition
to the correct exponent.\cite{WL_prefactor_footnote}

The final result for the nonanalytic contribution to the diffusion coefficient, and the corresponding LTT
of the velocity autocorrelation function, is
\bse
\label{eqs:4.13}
\bea
\frac{\Delta D(z)}{D_0} &=& - (n_i a^{d-1}\lambda)^{d-1} A_d \,
\nonumber\\
&& \hskip -40pt  \times
(-i z \tau)^{(d-2)/2}  \times \begin{cases} 1 & \text{for}\quad d=3\\
                                                                                                                        \ln(i z \tau) & \text{for}\quad d=2
                                                                                                 \end{cases}  \ , \qquad\quad
\label{eq:4.13a}
\eea
with
\be
A_d = d^{d/2}/2^d\pi^2\ ,
\label{eq:4.13b}
\ee
\ese
and (see Appendix~\ref{app:A})
\bse
\label{eqs:4.14}
\be
\lim_{t\to\infty}\langle {\bm v}(t)\cdot{\bm v}(t=0)\rangle = - v^2\,(n_i a^{d-1}\lambda)^{d-1} B_d \,(t/\tau)^{-d/2}\ ,
\label{eq:4.14a}
\ee
with
\be
B_d = d^{d/2}/2^{2(d-1)} \pi^{(d+2)/2}\ .
\label{eq:4.14b}
\ee
\ese

Equations~(\ref{eqs:4.13}, \ref{eqs:4.14}) represent the leading LTT in a QM Lorentz model.
It should be stressed that this result is valid only for non-interacting electrons in the presence
of time-reversal invariance. Breaking the latter, e.g., via magnetic impurities or a magnetic
field, gives the vertex $\Gamma^{\text{WL}}$ a mass and cuts off the singularity in
Eq.~(\ref{eq:4.13a}).\cite{Lee_Ramakrishnan_1985}

\section{Discussion}
\label{sec:V}

In summary, we have provided a unified, and mostly elementary, derivation of the long-time tails of the
velocity autocorrelation function in both classical and quantum Lorentz models. The results are as
summarized in Eqs.~(\ref{eqs:1.1}) and (\ref{eqs:1.2}), and our method also produces the exact
prefactors. In the remainder of this section we discuss various aspects of these results and the
underlying physics that were not covered in detail in the main text. 

\begin{enumerate}[wide,labelindent=0pt]
\item We start with a brief additional comment on the disorder expansion parameter in the
              classical and quantum cases, respectively. The length scale associated with the
              disorder is the mean-free path $\ell \propto 1/n_i a^{d-1}$, with $n_i$ the scatterer
              density and $a$ the scattering length. This length scale should be compared to the
              relevant length scale associated with the scattered particle. In a classical Lorentz
              model, there is only one such length scale, namely $a$, and hence the small
              parameter for the disorder expansion is $a/\ell \propto n_i a^d$. In a quantum
              system, there is a second length scale associated with the particle, namely, the
              de Broglie wavelength $\lambda$. For low temperatures, $\lambda\gg a$, and
              the leading expansion parameter is $\lambda/\ell \propto n_i a^{d-1}\lambda$.
              As stated in Eq.~(\ref{eq:1.3}), this parameter can also be written as $1/\kF\ell = \hbar/2\epsilonF\tau$,
              with $\kF$ and $\epsilonF$ the Fermi wave number and energy, respectively,
              and $\tau$ the mean-free time.
\item As we mentioned in Sec.~\ref{subsec:III.B}, the semi-classical LTT that naturally results
              from a random potential with a quantum-mechanical scattering cross section is
              not the leading LTT in a quantum Lorentz model, and its counterpart with
              a classical scattering cross section is not the leading one in a classical model either. This
              contribution is nevertheless very important for both historical and technical reasons. 
              First, a random scattering potential was the model setup that originally led to the concept
              of Anderson localization.\cite{Anderson_1958} Second, this contribution is what naturally
              emerges in any approach that starts with diffusing electrons, and the leading
              weak-localization LTT can be derived from it by a more careful analysis of the diffusive
              propagator. 
              
              Interestingly, the prefactor of the semi-classical term contains a high power of the
              coupling constant, as was mentioned after Eqs.~(\ref{eqs:3.17}). The modification that
              leads to the weak-localization LTT affects both the prefactor and the LTT exponent,
              see Eqs.~(\ref{eq:D.1}) and (\ref{eq:D.2}). 
              
\item The fact that the crossed-ladder diagrams in Fig.~\ref{fig:3} also represent the diffusion 
              propagator, as do the ladder diagrams in Fig.~\ref{fig:2}, but at wave-vector argument 
              ${\bm k}+{\bm p}$ rather than ${\bm q}$, reflects an exact symmetry of the 
              vertex function $\Gamma_{{\bm k}{\bm p}}$ and the phase-space correlation function 
              $\Phi_{{\bm k}{\bm p}}$, valid for non-interacting electrons in the presence of time-reversal 
              invariance, that was first noticed by Prelov\v{s}ek,\cite{Prelovsek_1981}
             \bse
             \label{eqs:5.1}
             \bea
             \Gamma_{{\bm k}{\bm p}}({\bm q},z) &=& \Gamma_{({\bm k}-{\bm p}+{\bm q})/2,({\bm p}-{\bm k}+{\bm q})/2}({\bm k}+{\bm p},z)\ ,\qquad
             \label{eq:5.1a}\\
             \Phi_{{\bm k}{\bm p}}({\bm q},z) &=& \Phi_{({\bm k}-{\bm p}+{\bm q})/2,({\bm p}-{\bm k}+{\bm q})/2}({\bm k}+{\bm p},z)\ ,
             \label{eq:5.1b}
             \eea
             \ese
             and used in Ref.~\onlinecite{Belitz_Gold_Goetze_1981} to derive the WL LTT from the semi-classical one. 
             The importance of time-reversal invariance is obvious in the diagrammatic representation:
             The crossed ladder can be obtained from the ladder by reversing the direction of the
             lower Green function, which is permissible only if time-reversal invariance holds. 

\item Even in the absence of time-reversal invariance there exists a LTT that is stronger
              than the semi-classical one represented by Eqs.~(\ref{eqs:3.16}), (\ref{eqs:3.17}); 
              the relevant model is the unitary matrix model studied by Wegner.\cite{Wegner_1989} 
              This leads to $\Delta D(z) \propto z^{d-2}$,\cite{Belitz_Yang_1993} which corresponds
              to a LTT of the velocity autocorrelation function proportional to $1/t^{d-1}$. The zero 
              frequency exponent in $d=2$ indicates a logarithm as in Eq.~(\ref{eq:4.13a}), which 
              in time space corresponds to a $1/t$ LTT. In the field-theoretic treatments of 
              Refs.~\onlinecite{Wegner_1989, Belitz_Yang_1993} this LTT appears at two-loop order. 
              It is unclear how, or if, this is related to the hydrodynamic arguments used in the present
              paper that capture one-loop effects.   
              
\item The LTTs are perturbative results, and it is not clear {\it a priori} what they signify for the
             fate of the phase in which they occur if the coupling constant is increased or/and the
             dimensionality is decreased. The LTTs we have derived all have in common that the
             diffusion coefficient is an increasing function of the frequency, so the static diffusion
             coefficient is decreased compared to the Boltzmann result. With increasing coupling
             constant the diffusive phase is ultimately destroyed and gives way to a phase where
             diffusion is blocked. This is true for the classical Lorentz model in both $d=2$ and 
             $d=3$,\cite{Alder_Alley_1978, Hoefling_Franosch_Frey_2006} as well as for the 
             quantum Lorentz model in $d=3$, as was realized by Anderson.\cite{Anderson_1958} In
             the quantum case in $d=2$ the LTT is strong enough to destroy the diffusive phase at
             arbitrarily small values of the coupling constant; this is the phenomenon of weak localization. 
             These phase transitions from a diffusive phase to a non-diffusive one can be described
             by making the LTT calculations self-consistent, i.e., replacing the bare diffusion
             coefficient under the wave-number integral by the physical one.\cite{Goetze_1979,
             Goetze_Leutheusser_Yip_1981}. 
             
             However, there are systems with an equally strong LTT with the opposite sign, which
             is sometimes referred to as `weak anti-localization', so the static diffusion coefficient  
             {\em increases} with increasing coupling constant.\cite{Evers_Mirlin_2008} With
             increasing coupling constant the static diffusion coefficient then goes through
             a maximum and eventually vanishes at a critical value of the coupling constant. 
             By contrast, the LTT in a $2-d$ classical fluid, which also has the `anti-localizing'
             sign, signalizes the breakdown of linearized hydrodynamics.\cite{Forster_Nelson_Stephen_1977}   
             
\item In various places we have neglected contributions to the generalized diffusion equation
             that are of $O(k^4)$ rather than $k^2$, see, e.g., Eq.~(\ref{eq:B.8}). These are
             renormalizations of a super-Burnett coefficient rather than the diffusion coefficient. 
             These higher order transport coefficients display increasingly strong LTTs. This signifies a 
              breakdown of the gradient expansion for the hydrodynamic modes. Indeed, the hydrodynamic
              frequencies in a classical fluid have nonanalytic dispersion relations with infinitely many
              terms between $k^2$ and $k^4$.\cite{Ernst_Dorfman_1975}

\item The quantum Lorentz model we have discussed is a model for non-interacting electrons.
            An electron-electron interaction leads to a host of new effects; for reviews, see 
            Refs.~\onlinecite{Altshuler_Aronov_1985, Belitz_Kirkpatrick_1994}. One interesting
            effect is that the long-time tail for the diffusion coefficient is as strong as the
            weak-localization one even in the absence of time-reversal invariance.\cite{Altshuler_Aronov_Lee_1980}
            Structurally, the interaction couples pairs of diffusive propagators ${\cal D}_{1,2}$,
            which leads to contributions of the form\cite{Belitz_Kirkpatrick_1994}
            \be
            \Delta D(z) \propto \frac{1}{V}\sum_{\bm q} {\bm q}^2\,{\cal D}_1({\bm q},z)\,{\cal D}_2({\bm q},z)    
                              \propto z^{(d-2)/2}\ ,
            \label{eq:5.2}
            \ee    
            with the zero exponent for $d=2$ to be interpreted as a logarithm.  Some of the diffusive modes
            involved are not hydrodynamic in nature, and it is not obvious whether, or how, the
            arguments employed in this paper can be adapted to describe these effects.    

\end{enumerate}

\appendix
\section{Low-Frequency Asymptotics and Long-Time Tails}
\label{app:A}

In this appendix we briefly recall how nonanalyticities in frequency space correspond to non-exponential
decays in time space. For details and proofs see Refs.~\onlinecite{Lighthill_1958} and \onlinecite{Feller_1970}.

\subsection{Fourier transforms and Laplace transforms}
\label{app:A.1}

Consider a time-dependent function $f(t)$ that for $t\to\pm\infty$ does not grow faster than some power
and define a Fourier transform
\bse
\label{eqs:A.1}
\be
{\hat f}(\omega) = \int_{-\infty}^{\infty} dt\,e^{i\omega t} f(t)\ ,
\label{eq:A.1a}
\ee
which makes the Fourier backtransform
\be
f(t) =\int_{-\infty}^{\infty} \frac{d\omega}{2\pi}\,e^{-i\omega t} {\hat f}(\omega)\ ,
\label{eq:A.1b}
\ee
\ese
and a Laplace transform
\be
F(z) = \pm\int_{-\infty}^{\infty}  dt\,\Theta(\pm t)\,e^{izt} f(t)\qquad (\pm\,\text{for}\,\text{Im}(z) \genfrac{}{}{0pt}{2}{>}{<} 0)
\label{eq:A.2}
\ee
with $z$ the complex frequency. $F(z)$ has a spectral representation
\bse
\label{eqs:A.3}
\be
F(z) = \int_{-\infty}^{\infty}  \frac{d\omega}{\pi}\,\frac{F''(\omega)}{\omega - z}
\label{eq:A.3a}
\ee
where the spectral density $F''$ of $F$ is given by
\be
F''(\omega) = \frac{1}{2i} \left[F(\omega + i0) - F(\omega - i0)\right] = \frac{1}{2i}\,{\hat f}(\omega)\ .
\label{eq:A.3b}
\ee
\ese

\subsection{Tauberian theorems}
\label{app:A.2}

The results that are known as Tauberian theorems (or Abelian theorems, depending on 
whether they relate $F(z)$ to $f(t)$ or vice versa) state that the analytic properties of
$F$ at $z=0$ determine the asymptotic behavior of $f$ for $\vert t\vert\to\infty$, and
vice versa. Specfically, if $F$ is holomorphic in a neighborhood of $z=0$, then $f$
decays exponentially for $\vert t\vert\to\infty$. Whereas, a nonanalyticity of $F$ at
$z=0$ implies a non-exponential asymptotic behavior of $f$, and vice versa. 
An equivalent statement is that a nonanalyticity of the spectral density $F''$ at
$\omega = 0$ implies non-exponential decay of $f$, and vice versa. Some useful 
examples are as follows.

\smallskip\par\noindent
{\em Example 1:} Non-integer powers. Let
\bse
\label{eqs:A.4}
\be
F(z) = z^{\alpha}\, 
\label{eq:A.4a}
\ee
with $\alpha > -1$ real and non-integer, plus terms that are holomorphic in a neighborhood of $z=0$. The spectral density is
\be
F''(\omega\to 0) = \sin(\pi\alpha)\,\Theta(-\omega)\,\vert\omega\vert^{\alpha}\ ,
\label{eq:A.4b}
\ee
with $\Theta$ the step function, and the asymptotic behavior of $f$ is
\be
f(t\to\pm\infty) = \frac{-1}{\pi}\,\sin(\pi\alpha)\,e^{i(\pi/2)\alpha\, \sgn t}\, \Gamma(\alpha+1)\,\frac{\sgn t}{\vert t\vert^{\alpha+1}}\ ,
\label{eq:A.4c}
\ee
\ese
with $\Gamma$ the Gamma function. 

For our purposes we are interested in a version of this example that is antisymmetric in $z$, namely (see Eq.~(\ref{eq:2.8c})
\bse
\label{eqs:A.5}
\be
F(z) = s(z) \left(-i z s(z)\right)^{\alpha}
\label{eq:A5a}
\ee
which has a spectral density
\be
F''(\omega) = -i\,\cos(\pi\alpha/2)\,\vert\omega\vert^{\alpha}\ .
\label{eq:A.5b}
\ee
The asymptotics of the underlying function $f$ are
\be
f(t\to\pm\infty) = \frac{-1}{\pi}\,\sin(\pi\alpha)\,\Gamma(\alpha+1)\,\frac{1}{\vert t\vert^{\alpha+1}}\ .
\label{eq:A.5c}
\ee
\ese
This example covers the LTTs in $d=3$ that were discussed in the main text: For $\alpha = 3/2$
it shows how Eqs.~(\ref{eq:3.10b}) and (\ref{eqs:3.17}) follow from Eqs.~(\ref{eqs:3.8}) and
(\ref{eqs:3.16}), respectively. For $\alpha = 1/2$ it shows how Eqs.~(\ref{eqs:4.14}) follow from
Eqs.~(\ref{eqs:4.13}). 

We note in passing that the function 
$$G(z) =  i\left[z^{\alpha} - (-z)^{\alpha}\right]\ ,$$
which has the same symmetry properties as $F(z)$, has a spectrum $G''(\omega)$ that is 
proportional to $F''(\omega)$ and thus contains the same information about the LTT. We also 
note that, within the theory of generalized functions, the Fourier transform of Eq.~(\ref{eq:A.5c}) 
exists for all non-integer $\alpha$, and Fourier's inversion theorem holds.\cite{Lighthill_1958}



\smallskip\par\noindent
{\em Example 2:} Integer powers multiplied by logarithms. Let
\bse
\label{eqs:A.6}
\be
F(z) = (-z)^n \ln z
\label{eq:A.6a}
\ee
plus holomorphic parts, with $n$ an integer $> -1$. Then the spectral density is
\be
F''(\omega\to 0) = \pi \vert\omega\vert^n \Theta(-\omega)
\label{eq:A.6b}
\ee
and
\be
f(t\to\pm\infty) = - i^n\,n!\,\frac{1}{t^{n+1}}\ .
\label{eq:A.6c}
\ee
\ese

For the LTTs in $d=2$ we need the following modification of Eq.~(\ref{eq:A.6a}) (see Eq.~(\ref{eq:2.8c})):
\be
F(z) = s(z)\,\left(-iz s(z)\right)^n \ln\left(-i z s(z)\right)
\label{eq:A.7}
\ee
Now we need to distinguish between odd and even $n$. 
\medskip\par\noindent
{\it Example 2a):} Let $n = 2m+1$, with $m\geq 0$ integer. Then
\bse
\label{eqs:A.8}
\be
F''(\omega) = (-)^m\,\frac{i\pi}{2}\,\vert\omega\vert^{2m+1}
\label{eq:A.8a}
\ee
and the asymptotic behavior of the underlying time-dependent function is
\be
f(t\to\pm\infty) = (2m+1)!/t^{2(m+1)}
\label{eq:A.8b}
\ee
\ese
For $m=0$ this example shows how Eqs.~(\ref{eqs:3.10}) follow from Eqs.~(\ref{eqs:3.8}) in $d=2$.
\medskip\par\noindent
{\it Example 2b):} Let $n = 2m$, with $m\geq 0$ integer. Then
\bse
\label{eqs:A.9}
\be
F''(\omega) = i (-)^{m+1}\,\omega^{2m}\,\ln \vert\omega\vert\ ,
\label{eq:A.9a}
\ee
and
\be
f(t\to\pm\infty) = -(2m)!/\vert t\vert^{2m+1}
\label{eq:A.9b}
\ee
\ese
For $m=0$ this example shows how Eqs.~(\ref{eqs:4.14}) follow from Eqs.~(\ref{eqs:4.13}) in $d=2$. 

Note that the simple antisymmetrization 
$$G(z) = i\left[ (-z)^n \ln z - z^n \ln(-z)\right]$$
yields the same information if $n$ is odd, but not if $n$ is even. In the latter case, the spectral density $G''$ 
does not contain the $\ln \vert\omega\vert$ term that is necessary for obtaining the LTT. Finally the Fourier 
transforms of Eqs.~(\ref{eq:A.8b}) and (\ref{eq:A.9b}) again exist within the theory of generalized functions and 
correctly yield ${\hat f}(\omega)$ (in the case of Eq.~(\ref{eq:A.9b}) with an additional term that replaces 
$\ln \vert\omega\vert$ with an arbitrary constant), so Fourier's inversion theorem holds.\cite{Lighthill_1958}

We add three remarks. 
(1) Different authors use different conventions for defining the Fourier and Laplace transforms. 
(2) The Tauberian theorems as asymptotic statements can be proven within classical analysis. 
The task is made easier by the notion of generalized functions,\cite{Lighthill_1958} which allows
for the definition of Fourier transforms that do not exist within classical analysis. 
(3) Knowledge of $F(z)$ on the positive imaginary axis, $z = i\lambda$ with $\lambda>0$, 
suffices for obtaining the asymptotic behavior of $f(t)$ for $t>0$, see Ref.~\onlinecite{Feller_1970}
Ch.~13 Sec.~2 Examples~(b).

\medskip
\section{Alternative ways of calculating the long-time tails}
\label{app:B}

In this appendix we show how to (1) calculate the density-density time-correlation function in the
presence of a fluctuating potential without making use of time-reversal invariance, and (2) how
to obtain the renormalized diffusion coefficient by considering the decay of an externally
created density perturbation. For either goal one needs to address a technical issue regarding
the disorder average over a random potential, so we discuss this first. 

\subsection{Ensemble average for a fixed random potential}
\label{app:B.1}

Averages of quantities that are linear in the random potential $U({\bm q})$, such as the second
term on the right-hand side of Eq.~(\ref{eq:3.12}), do not vanish since the potential enters the
Hamiltonian and hence the ensemble average $\langle\ldots\rangle_{\text{ens}}$. Consider
the equal-time density correlation function for a fixed realization of the random potential, so
the correlation function involves an ensemble average only. Let ${\bm r}$ be the position of the
particle at time $t$. Then classically, $\delta n({\bm x},t) = \delta({\bm x}-{\bm r})$. For a given
realization of the random potential, the ensemble average consists of averaging over all
possible positions ${\bm r}$ with a Boltzmann weight:\cite{classical_Stat_Mech_footnote}
\bse
\label{eqs:B.1}
\be
\langle\delta n({\bm k},t)\,\delta n({\bm p},t)\rangle_{\text{ens}} = \frac{1}{Z}\int_V d{\bm r}\,e^{-U({\bm r})/T}\,e^{-i({\bm k}+{\bm p})\cdot{\bm r}}\ .
\label{eq:B.1a}
\ee
where
\be
Z = \int_V d{\bm r}\,e^{-U({\bm r})/T}
\label{eq:B.1b}
\ee
\ese
and we use units such that $\kB=1$.
By the equipartition theorem, the temperature is related to the particle's kinetic energy by
\be
T = mv^2/d = m D_0/\tau\ .
\label{eq:B.2}
\ee
The spatial average over a particular realization of $U({\bm r})$ must be equal to the
average over the ensemble of random potentials at a fixed position, so Eq.~(\ref{eq:2.17a})
implies $\int_V d{\bm r}\,U({\bm r}) = 0$. Expanding in powers of $U$ we thus have
\be
\langle\delta n({\bm k},t)\,\delta n({\bm p},t)\rangle_{\text{ens}} = \delta_{{\bm k},-{\bm p}} - \frac{1}{TV}\,U({\bm k}+{\bm p}) + O(U^2)\ .
\label{eq:B.3}
\ee
This implies
\bea
\langle U({\bm k}-{\bm q})\,\delta n({\bm q},t=0)\,\delta n(-{\bm k},t=0)\rangle &=& \frac{-\tau}{mD_0}\,{\cal U}^2({\bm k}-{\bm q})
\nonumber\\
&& \hskip 0pt + O(U^4)\ ,
\label{eq:B.4}
\eea
where we have used Eqs.~(\ref{eq:2.17b}) and (\ref{eq:B.2}). We will make use of this expression in the next subsection. 

Now consider the case of a density perturbation that is externally created at a position ${\bm r}_0$. Then
$\delta n({\bm x},t) = \delta({\bm x}-{\bm r})\,V\,\delta({\bm r}-{\bm r}_0)$. For a given realization of the
random potential we have
\bse
\label{eqs:B.5}
\bea
\langle \delta n({\bm k},t=0)\rangle_{\text{ens}} &=& \frac{V}{Z} \int_V d{\bm r}\,e^{-U({\bm r})/T}\,e^{-i{\bm k}\cdot{\bm r}_0} \,\delta({\bm r}-{\bm r}_0)
\nonumber\\
&=& e^{-U({\bm r}_0)/T}\,e^{-i{\bm k}\cdot{\rm r}_0} \left[1 + O(U^2)\right]\ .
\label{eq:B.5a}
\eea
and a disorder average yields
\be
\langle\delta n({\bm k},t=0)\rangle = e^{-i{\bm k}\cdot{\bm r}_0} + O(U^2)\ .
\label{eq:B.5b}
\ee
\ese
This implies
\bea
\langle U({\bm k}-{\bm q})\,\delta n({\bm q},t=0)\rangle &=& \frac{-\tau}{m D_0}\,{\cal U}^2({\bm k}-{\bm p})\,\langle\delta n({\bm k},t=0)\rangle
\nonumber\\
&& \hskip 40pt  + O(U^4)\ ,
\label{eq:B.6}
\eea
which we will use in Appendix~\ref{app:B.4.b}.

\subsection{The density-density correlation function in the absence of time-reversal symmetry}
\label{app:B.2}

Here we show that time reversal invariance, which we used in the last line of Eq.~(\ref{eq:3.11a}), is not
necessary for obtaining the result expressed in Eqs.~(\ref{eq:3.14}). 

We start with the first line in Eq.~(\ref{eq:3.11a}) and follow the same procedure that led to Eq.~(\ref{eq:3.12}).
We obtain
\begin{widetext}
\bea
(-iz + D_0 k^2) S_{nn}({\bm k},z) &=& S_{nn}^{\,0} - \frac{\tau}{m}\,\frac{1}{V} \sum_{\bm q} {\bm k}\cdot({\bm k}-{\bm q})\,{\frak D}({\bm q},z)\,
        \langle U({\bm k}-{\bm q})\,\delta n({\bm q},t=0)\,\delta n(-{\bm k},t=0)\rangle
\nonumber\\
&& \hskip 19pt + \frac{\tau^2}{m^2}\,\frac{1}{V^2} \sum_{{\bm q},{\bm q}'} \left({\bm k}\cdot({\bm k}-{\bm q})\right)\,\left({\bm q}\cdot({\bm q}-{\bm q}')\right)\,
       {\frak D}({\bm q},z)\,{\frak D}({\bm q}',z)\,
\nonumber\\
&& \hskip 85pt \times       \langle U({\bm k}-{\bm q})\,U({\bm q}-{\bm q}')\,\delta n({\bm q}',t=0)\,\delta n(-{\bm k},t=0)\rangle 
\label{eq:B.7}
\eea
Factorizing the last term on the right-hand side yields Eq.~(\ref{eq:3.14}) for the correction $\Delta D(z)$ to the diffusion coefficient,
and in addition another contribution equal to $d\,\Delta D(z)$. We now combine this with what results from the second term on the
right-hand side if we use Eq.~(\ref{eq:B.4}). We then obtain a contribution to $\Delta D$, in addition to Eq.~(\ref{eq:3.14}), given by
\bea
\Delta D^{(1)}({\bm k},z) &=& -\left(\frac{\tau}{m}\right)^2 \frac{1}{V} \sum_{\bm q} {\cal U}^2({\bm q})\,\left[q^2
                                          + {\frak D}^{-1}({\bm k},z)\,\frac{1}{D_0}\right]\,\,{\frak D}({\bm q},z)
\nonumber\\
&=& - \frac{\tau^2}{D_0 m^2}   \frac{1}{V} \sum_{\bm q} {\cal U}^2({\bm q})\,\left[1 + D_0\, k^2\, {\frak D}({\bm q},z)\right]\ .                                         
\label{eq:B.8}
\eea
\end{widetext}
We see that the additional contributions to $\Delta D$ amount to a term that does not involve an integral over a diffusion
pole, and hence does not contribute to the LTT, and another term that contributes at higher order in $k$, and thus 
represents a renormalization of a super-Burnett coefficient rather than the diffusion coefficient. 

This shows that the result for the density time-correlation function, and the renormalized diffusion coefficient, is the
same whether or not we assume time reversal invariance.

\subsection{Decay of a density perturbation}
\label{app:B.3}

In this appendix we show how to obtain the LTTs by considering the decay of an externally created
macroscopic density perturbation.  As expected, the renormalization of the average $\langle\delta n\rangle$ 
is consistent with that of the two-point correlation function $\langle\delta n\,\delta n\rangle$.

\subsubsection{Fluctuating diffusion coefficient}
\label{app:B.4.a}

Consider Eq.~(\ref{eq:2.6c}) and perform an average. The equation then describes
the decay of a density perturbation from an initial condition $\langle\delta n({\bm k},t=0)\rangle$.
By iterating the equation once before averaging we obtain
\begin{widetext}
\bea
(-iz + D_0 k^2)\langle\delta n({\bm k},z)\rangle &=& \langle\delta n({\bm k},t=0) \rangle
 - \frac{1}{V}\sum_{\bm q}({\bm k}\cdot{\bm q})\,{\frak D}({\bm q},z) \langle\delta D({\bm k}-{\bm q})\,\delta n({\bm q},t=0)\rangle
\nonumber\\
&& \hskip 62pt + \frac{1}{V^2}\sum_{{\bm q},{\bm q}'}({\bm k}\cdot{\bm q})({\bm q}\cdot{\bm q}') {\frak D}({\bm q},z) \langle \delta D({\bm k}-{\bm q})\,\delta D({\bm q}-{\bm q}')\,\delta n({\bm q}',z)\rangle\ .
\label{eq:B.9}
\eea
Upon factorizing the averages as in Sec.~\ref{subsubsec:III.A.1} the second term on the right-hand side vanishes,
and we obtain
\be
(-iz + D_0 k^2)\langle\delta n({\bm k},z)\rangle = \langle\delta n({\bm k},t=0)\rangle
+ \frac{1}{V}\sum_{\bm q}({\bm k}\cdot{\bm q})^2 \Delta({\bm k}-{\bm q})\,{\frak D}({\bm q},z)\,\langle\delta n({\bm k},z)\rangle\ .\qquad
\label{eq:B.10}
\ee
The equation again retains its diffusive form, the disorder renormalizes the diffusion coefficient,
and the correction $\Delta D(z)$ to the latter is given by Eq.~(\ref{eq:3.6}).

\subsubsection{Random potential}
\label{app:B.4.b}

Now iterate Eq.~(\ref{eq:2.16c}) once and average. This yields
\bea
(-iz + D_0 k^2)\langle\delta n({\bm k},z)\rangle &=& \langle\delta n({\bm k},t=0) \rangle
 - \frac{\tau}{m}\,\frac{1}{V}\sum_{\bm q}\left({\bm k}\cdot({\bm k}-{\bm q})\right)\,{\frak D}({\bm q},z) \langle{\cal U}({\bm k}-{\bm q})\,\delta n({\bm q},t=0)\rangle
\nonumber\\
&& \hskip 0pt + \frac{\tau^2}{m^2}\,\frac{1}{V^2}\sum_{{\bm q},{\bm q}'}({\bm k}\cdot{\bm q})\left(({\bm k}-{\bm q})\cdot{\bm q}'\right) {\frak D}({\bm k}-{\bm q},z) \langle {\cal U}({\bm q})\,{\cal U}({\bm q}')\,\delta n({\bm k}-{\bm q}-{\bm q}',z)\rangle\ .
\label{eq:B.11}
\eea
\end{widetext}
Now we use Eq.~(\ref{eq:B.6}) in the second term on the right-hand side, and factorize the average in the third term. 
This gives the same result as in Appendix~\ref{app:B.2}, viz., $\Delta D(z)$ from Eq.~(\ref{eq:3.14}) plus the additional
contribution $\Delta D^{(1)}$ from Eq.~(\ref{eq:B.8}) which does not contribute to the LTT of the diffusion coefficient.

\section{Relation to mode-coupling theory}
\label{app:C}

The result expressed in Eqs.~(\ref{eq:3.14}), (\ref{eq:3.15}) was first derived by G{\"o}tze\cite{Goetze_1979}
in the framework of a mode-coupling theory.\cite{mode_coupling_footnote} Historically, this type of theory goes back to
Kadanoff and Swift\cite{Kadanoff_Swift_1968} and Kawasaki\cite{Kawasaki_1970} who used it to study transport coefficients near
critical points. It was used to derive the LTT in a classical fluid in Refs.~\onlinecite{Ernst_Hauge_van_Leeuwen_1971,
Ernst_Hauge_van_Leeuwen_1976a, Ernst_Hauge_van_Leeuwen_1976b}. Here we briefly
describe the connection to the procedure in Sec.~\ref{subsec:III.B}. To avoid misunderstandings
we mention that the main goal of Ref.~\onlinecite{Goetze_1979} was to derive a theory
for an Anderson transition by making the mode-coupling theory self-consistent, whereas
in this paper we focus entirely on the LTT in the diffusive phase. 

The correlation function considered in Ref.~\onlinecite{Goetze_1979} is the density-density
Kubo function $\Phi_{nn}({\bm k},z)$, which is given in terms of the density-density
susceptibility $\chi_{nn}$ by (see Eq.~(\ref{eq:4.3b}))
\be
\Phi_{nn}({\bm k},z) = \frac{1}{z}\,\left[ \chi_{nn}({\bm k},z) - \chi_{nn}({\bm k},z=0)\right]\ .
\label{eq:C.1}
\ee
The spectral density of the susceptibility, $\chi_{nn}''({\bm k},\omega)  = \lim_{\epsilon\to 0}\text{Im}\,\chi_{nn}({\bm k},\omega + i\epsilon)$,
is related to the correlation function $S_{nn}$ by
\be
S_{nn}^{\text{sym}}({\bm k},\omega) = \hbar\,\coth(\hbar\omega/2T)\,\chi_{nn}''({\bm k},\omega) \ .
\label{eq:C.2}
\ee
$S_{nn}^{\text{sym}}$ is the symmetrized version of $S_{nn}$, i.e., the anticommutator density correlation function.
In the classical limit this turns into Eq.~(\ref{eq:4.2a}).

$\Phi_{nn}$ can be written as a continued fraction in a scheme proposed by Zwanzig\cite{Zwanzig_1961} and
Mori,\cite{Mori_1965}
\bse
\label{eqs:C.3}
\be
\Phi_{nn}({\bm k},z) = \frac{-\chi_{nn}({\bm k},z=0)}{z+ k^2 K({\bm k},z)}\ ,
\label{eq:C.3a}
\ee
with
\be
K({\bm k},z) = \frac{-\vF^2/d}{z + M({\bm k},z)}
\label{eq:C.3b}
\ee
\ese
the current-current correlation function and $M({\bm k},z)$ the current relaxation kernel or
memory function. The latter is given by a projected, or reduced, correlation function of the
time derivative of the current, and Eqs.~(\ref{eqs:C.3}) are formally exact.

To evaluate $M$, the mode-coupling theory of Ref.~\onlinecite{Goetze_1979} makes two approximations
that go beyond Drude theory. First, it neglects the wave-vector dependence of the kernel, i.e., it replaces
$M({\bm k},z)$ by $M(z) \equiv M({\bm k}=0,z)$. This makes $M(z)$ proportional to the Kubo
correlation function of the wave-number convolution of the gradient of $U$ and $\delta n$ that appears
in Eqs.~(\ref{eqs:3.11}). Second, it factorizes the resulting autocorrelation function. The former approximation
is equivalent to neglecting terms of $O(k^4)$ (see the remark
after Eq.~(\ref{eq:B.8})), and the latter is the same factorization that led from Eq.~(\ref{eq:3.12})
to (\ref{eq:3.13}). This makes $M(z)$ the product of $\{U\,U\}_{\text{dis}}$ and an integral over
the density-density Kubo correlation function:
\be
M(z) =  \frac{1}{2\pi\hbar\NF^2\kF^2\tau}\,\frac{1}{V}\sum_{\bm q} q^2\,\Phi_{nn}({\bm q},z)
\label{eq:C.4}
\ee
with $\kF$ the Fermi wave number. {$\Phi_{nn}$ behaves qualitatively differently for
wave numbers $q$ greater and smaller, respectively, than the inverse mean-free path
$1/\ell$. For $q \agt 1/\ell$ it is given by the free-electron correlation function. This part
of the ${\bm q}$-integral}
yields the Drude contribution to $M(z)$, which equals $i/\tau$.\cite{Goetze_Woefle_1972} 
With this result for $M(z)$, and in the limit of small
wave numbers and frequencies, Eq.~(\ref{eq:C.3a})  becomes proportional to the bare diffusion
propagator from Eq.~(\ref{eq:2.7}),
\be
\Phi_{nn}({\bm k},z) \approx i\NF\,{\frak D}({\bm k},z)\ .
\label{eq:C.5}
\ee
{This form of $\Phi_{nn}$ must be used for the small-$q$ part of the integral in Eq.~(\ref{eq:C.4}).
We thus have}
\bse
\label{eqs:C.6}
\be
M(z) = i/\tau + m(z)\ ,
\label{eq:C.6a}
\ee
with $m(z)$ determined by the diffusive part of the density-density correlation, i.e., by
Eq.~(\ref{eq:C.5}) used in Eq.~(\ref{eq:C.4}). This yields\cite{Goetze_1979, Goetze_1981}
\be
m(z) = \frac{i}{2\pi\hbar\NF\kF^2\tau}\,\frac{1}{V}{\sum_{\bm q}}^{\prime} q^2\,{\frak D}({\bm q},z)\ ,
\label{eq:C.6b}
\ee
\ese
where the prime on the sum over ${\bm q}$ indicates that the integral
must be cut off at $q\approx 1/\ell$. For larger wave numbers $\Phi_{nn}$ crosses over to
its ballistic part whose effect is already contained in the Drude contribution to $M(z)$.

The correction to the diffusion coefficient is given in terms of $m(z)$ by
\be
{ \Delta D(z)/D_0 = i\tau m(z)\ . }
\label{eq:C.7}
\ee
Combining Eqs.~(\ref{eq:C.6b}) and (\ref{eq:C.7}) we 
obtain Eq.~(\ref{eq:3.15}). This establishes the equivalence of the hydrodynamic arguments
in Sec.~\ref{subsec:III.B} and the mode-coupling theory.

\section{A subleading quantum long-time tail}
\label{app:D}

The leading quantum LTT discussed in Sec.~\ref{sec:IV} is much stronger than the
one in Sec.~\ref{subsec:III.B}: The time decay is slower, $1/t^{d/2}$ vs. 
$1/t^{(d+2)/2}$, and the prefactor is larger, $(n_a a^{d-1}\lambda)^{d-1}$ vs.
$(n_a a^{d-1}\lambda)^{d+1}$, see Eqs.~(\ref{eqs:4.14}) and (\ref{eqs:3.17}). This
raises the question whether the same weak-localization mechanism in the context
of a fluctuating diffusion coefficient, i.e., the quantum version of the leading LTT in
a classical Lorentz model,  leads to a similar enhancement, which potentially could dominate
the standard weak-localization effect. This is not the case, which can be seen as follows.

The gradient squared in the integrand that determines $\Delta D(z)$, Eq.~(\ref{eq:4.4})
or (\ref{eq:3.15}) scales as
\be
q^2 \sim -iz/D_0 \propto -iz\tau/\vF^2\tau^2\ .
\label{eq:D.1}
\ee
Replacing this factor of $q^2$ by $2\kF^2$, which is what the weak-localization mechanism
effectively does, therefore multiplies the LTT by a factor proportional to
\be
\frac{1}{-iz\tau}\,\vF^2 \kF^2 \tau^2 \propto \frac{1}{-iz\tau}\,\frac{1}{(n_i a^{d-1}\lambda)^2}\ ,
\label{eq:D.2}
\ee
which leads to the standard weak-localization result, Eqs.~(\ref{eqs:4.13}, \ref{eqs:4.14}).

Now consider the mechanism due to a fluctuating diffusion coefficient. Here the relevant
integrand contains, in addition to the gradient squared,  a factor of $\Delta({\bm q})$, see Eq.~(\ref{eq:3.6}), which describes
the fluctuations of the diffusion coefficient, Eq.~(\ref{eq:2.10c}). The weak-localization mechanism
replaces $q^2$ by $2\kF^2$ as in the case of a fluctuating potential, but it also replaces
$\delta D(q=0)$ by $\delta D(q\approx \kF) \approx (\hbar/\epsilonF\tau)^2\,\delta D(q=0)$.
This leads to an additional factor of $(n_i a^2\lambda)^4$ from $\Delta(q\approx\kF)$. 
The factor that modifies the classical LTT thus is
\be
\frac{1}{-iz\tau}\,(n_i a^{d-1}\lambda)^2\ .
\label{eq:D.3}
\ee
This leads to the LTT given in Eq.~(\ref{eq:1.2b}), whose prefactor is smaller than the
standard weak-localization one by a factor of $(n_i a^{d-1}\lambda)^{3-d} (n_i a^d)^{d-1}$. 



\begin{thebibliography}{63}%
\makeatletter
\providecommand \@ifxundefined [1]{%
 \@ifx{#1\undefined}
}%
\providecommand \@ifnum [1]{%
 \ifnum #1\expandafter \@firstoftwo
 \else \expandafter \@secondoftwo
 \fi
}%
\providecommand \@ifx [1]{%
 \ifx #1\expandafter \@firstoftwo
 \else \expandafter \@secondoftwo
 \fi
}%
\providecommand \natexlab [1]{#1}%
\providecommand \enquote  [1]{``#1''}%
\providecommand \bibnamefont  [1]{#1}%
\providecommand \bibfnamefont [1]{#1}%
\providecommand \citenamefont [1]{#1}%
\providecommand \href@noop [0]{\@secondoftwo}%
\providecommand \href [0]{\begingroup \@sanitize@url \@href}%
\providecommand \@href[1]{\@@startlink{#1}\@@href}%
\providecommand \@@href[1]{\endgroup#1\@@endlink}%
\providecommand \@sanitize@url [0]{\catcode `\\12\catcode `\$12\catcode
  `\&12\catcode `\#12\catcode `\^12\catcode `\_12\catcode `\%12\relax}%
\providecommand \@@startlink[1]{}%
\providecommand \@@endlink[0]{}%
\providecommand \url  [0]{\begingroup\@sanitize@url \@url }%
\providecommand \@url [1]{\endgroup\@href {#1}{\urlprefix }}%
\providecommand \urlprefix  [0]{URL }%
\providecommand \Eprint [0]{\href }%
\providecommand \doibase [0]{http://dx.doi.org/}%
\providecommand \selectlanguage [0]{\@gobble}%
\providecommand \bibinfo  [0]{\@secondoftwo}%
\providecommand \bibfield  [0]{\@secondoftwo}%
\providecommand \translation [1]{[#1]}%
\providecommand \BibitemOpen [0]{}%
\providecommand \bibitemStop [0]{}%
\providecommand \bibitemNoStop [0]{.\EOS\space}%
\providecommand \EOS [0]{\spacefactor3000\relax}%
\providecommand \BibitemShut  [1]{\csname bibitem#1\endcsname}%
\let\auto@bib@innerbib\@empty
\bibitem [{\citenamefont {Alder}\ and\ \citenamefont
  {Wainwright}(1970)}]{Alder_Wainwright_1970}%
  \BibitemOpen
  \bibfield  {author} {\bibinfo {author} {\bibfnamefont {B.~J.}\ \bibnamefont
  {Alder}}\ and\ \bibinfo {author} {\bibfnamefont {T.~E.}\ \bibnamefont
  {Wainwright}},\ }\bibfield  {title} {\enquote {\bibinfo {title} {Decay of the
  velocity autocorrelation function},}\ }\href@noop {} {\bibfield  {journal}
  {\bibinfo  {journal} {Phys. Rev. A}\ }\textbf {\bibinfo {volume} {1}},\
  \bibinfo {pages} {18} (\bibinfo {year} {1970})}\BibitemShut {NoStop}%
\bibitem [{\citenamefont {Dorfman}\ and\ \citenamefont
  {Cohen}(1970)}]{Dorfman_Cohen_1970}%
  \BibitemOpen
  \bibfield  {author} {\bibinfo {author} {\bibfnamefont {J.~R.}\ \bibnamefont
  {Dorfman}}\ and\ \bibinfo {author} {\bibfnamefont {E.~G.~D.}\ \bibnamefont
  {Cohen}},\ }\bibfield  {title} {\enquote {\bibinfo {title} {Velocity
  correlation functions in two and three dimensions},}\ }\href@noop {}
  {\bibfield  {journal} {\bibinfo  {journal} {Phys. Rev. Lett.}\ }\textbf
  {\bibinfo {volume} {25}},\ \bibinfo {pages} {1257} (\bibinfo {year}
  {1970})}\BibitemShut {NoStop}%
\bibitem [{\citenamefont {Ernst}\ \emph {et~al.}(1970)\citenamefont {Ernst},
  \citenamefont {Hauge},\ and\ \citenamefont {van
  Leeuwen}}]{Ernst_Hauge_van_Leeuwen_1970}%
  \BibitemOpen
  \bibfield  {author} {\bibinfo {author} {\bibfnamefont {M.~H.}\ \bibnamefont
  {Ernst}}, \bibinfo {author} {\bibfnamefont {E.~H.}\ \bibnamefont {Hauge}}, \
  and\ \bibinfo {author} {\bibfnamefont {J.~M.~J.}\ \bibnamefont {van
  Leeuwen}},\ }\bibfield  {title} {\enquote {\bibinfo {title} {Asymptotic time
  behavior of correlation functions},}\ }\href@noop {} {\bibfield  {journal}
  {\bibinfo  {journal} {Phys. Rev. Lett.}\ }\textbf {\bibinfo {volume} {25}},\
  \bibinfo {pages} {1254} (\bibinfo {year} {1970})}\BibitemShut {NoStop}%
\bibitem [{\citenamefont {Ernst}\ \emph
  {et~al.}(1976{\natexlab{a}})\citenamefont {Ernst}, \citenamefont {Hauge},\
  and\ \citenamefont {van Leeuwen}}]{Ernst_Hauge_van_Leeuwen_1976a}%
  \BibitemOpen
  \bibfield  {author} {\bibinfo {author} {\bibfnamefont {M.~H.}\ \bibnamefont
  {Ernst}}, \bibinfo {author} {\bibfnamefont {E.~H.}\ \bibnamefont {Hauge}}, \
  and\ \bibinfo {author} {\bibfnamefont {J.~M.~J.}\ \bibnamefont {van
  Leeuwen}},\ }\bibfield  {title} {\enquote {\bibinfo {title} {Asymptotic time
  behavior of correlation functions. {II}. kinetic and potential terms},}\
  }\href@noop {} {\bibfield  {journal} {\bibinfo  {journal} {J. Stat. Phys.}\
  }\textbf {\bibinfo {volume} {15}},\ \bibinfo {pages} {7} (\bibinfo {year}
  {1976}{\natexlab{a}})}\BibitemShut {NoStop}%
\bibitem [{\citenamefont {Ernst}\ \emph
  {et~al.}(1976{\natexlab{b}})\citenamefont {Ernst}, \citenamefont {Hauge},\
  and\ \citenamefont {van Leeuwen}}]{Ernst_Hauge_van_Leeuwen_1976b}%
  \BibitemOpen
  \bibfield  {author} {\bibinfo {author} {\bibfnamefont {M.~H.}\ \bibnamefont
  {Ernst}}, \bibinfo {author} {\bibfnamefont {E.~H.}\ \bibnamefont {Hauge}}, \
  and\ \bibinfo {author} {\bibfnamefont {J.~M.~J.}\ \bibnamefont {van
  Leeuwen}},\ }\bibfield  {title} {\enquote {\bibinfo {title} {Asymptotic time
  behavior of correlation functions. {III}. local equilibrium and mode-coupling
  theory},}\ }\href@noop {} {\bibfield  {journal} {\bibinfo  {journal} {J.
  Stat. Phys.}\ }\textbf {\bibinfo {volume} {15}},\ \bibinfo {pages} {23}
  (\bibinfo {year} {1976}{\natexlab{b}})}\BibitemShut {NoStop}%
\bibitem [{\citenamefont {Ernst}\ and\ \citenamefont
  {Weyland}(1971)}]{Ernst_Weyland_1971}%
  \BibitemOpen
  \bibfield  {author} {\bibinfo {author} {\bibfnamefont {M.~H.}\ \bibnamefont
  {Ernst}}\ and\ \bibinfo {author} {\bibfnamefont {A.}~\bibnamefont
  {Weyland}},\ }\bibfield  {title} {\enquote {\bibinfo {title} {Long time
  behavior of the velocity auto-correlation function in a lorentz gas},}\
  }\href@noop {} {\bibfield  {journal} {\bibinfo  {journal} {Phys. Lett.}\
  }\textbf {\bibinfo {volume} {34A}},\ \bibinfo {pages} {39} (\bibinfo {year}
  {1971})}\BibitemShut {NoStop}%
\bibitem [{LTT()}]{LTT_sign_footnote}%
  \BibitemOpen
  \href@noop {} {}\bibinfo {note} {The sign can be attributed to the fact that
  the Lorentz gas has only one soft mode.\cite{Ernst_Weyland_1971} More
  precisely, the LTT of the shear viscosity in a classical fluid arises from
  the convective nonlinearity in the Navier-Stokes equations. It thus results
  from the coupling of two modes that carry one gradient each. By contrast, the
  LTT in a Lorentz model is due to the coupling of two modes that carry two
  gradients each, which accounts for the different sign. LTTs with the same
  sign as in the Lorentz gas also exist in fluids as a result of viscous
  terms.}\BibitemShut {Stop}%
\bibitem [{\citenamefont {Kovtun}\ and\ \citenamefont
  {Yaffe}(2003)}]{Kovtun_Yaffe_2003}%
  \BibitemOpen
  \bibfield  {author} {\bibinfo {author} {\bibfnamefont {P.}~\bibnamefont
  {Kovtun}}\ and\ \bibinfo {author} {\bibfnamefont {L.~G.}\ \bibnamefont
  {Yaffe}},\ }\bibfield  {title} {\enquote {\bibinfo {title} {Hydrodynamic
  fluctuations, long-time tails, and supersymmetry},}\ }\href@noop {}
  {\bibfield  {journal} {\bibinfo  {journal} {Phys. Rev. D}\ }\textbf {\bibinfo
  {volume} {68}},\ \bibinfo {pages} {025007} (\bibinfo {year}
  {2003})}\BibitemShut {NoStop}%
\bibitem [{\citenamefont {Caron-Huot}\ and\ \citenamefont
  {Saremi}(2010)}]{Caron-Huot_Saremi_2010}%
  \BibitemOpen
  \bibfield  {author} {\bibinfo {author} {\bibfnamefont {S.}~\bibnamefont
  {Caron-Huot}}\ and\ \bibinfo {author} {\bibfnamefont {O.}~\bibnamefont
  {Saremi}},\ }\bibfield  {title} {\enquote {\bibinfo {title} {Hydrodynamic
  long-time tails from anti de sitter space},}\ }\href@noop {} {\bibfield
  {journal} {\bibinfo  {journal} {J. High Energ. Phys.}\ }\textbf {\bibinfo
  {volume} {2010}},\ \bibinfo {pages} {13} (\bibinfo {year}
  {2010})}\BibitemShut {NoStop}%
\bibitem [{\citenamefont {Shukla}(2021)}]{Shukla_2021}%
  \BibitemOpen
  \bibfield  {author} {\bibinfo {author} {\bibfnamefont {A.}~\bibnamefont
  {Shukla}},\ }\bibfield  {title} {\enquote {\bibinfo {title} {Hydrodynamic
  fluctuations and long-time tails on an anisotropic background},}\ }\href@noop
  {} {\bibfield  {journal} {\bibinfo  {journal} {Nucl. Phys. B}\ }\textbf
  {\bibinfo {volume} {968}},\ \bibinfo {pages} {115442} (\bibinfo {year}
  {2021})}\BibitemShut {NoStop}%
\bibitem [{\citenamefont {Abbasi}\ and\ \citenamefont
  {Tabatabaei}(2020)}]{Abbasi_Tabatabaei_2020}%
  \BibitemOpen
  \bibfield  {author} {\bibinfo {author} {\bibfnamefont {N.}~\bibnamefont
  {Abbasi}}\ and\ \bibinfo {author} {\bibfnamefont {J.}~\bibnamefont
  {Tabatabaei}},\ }\bibfield  {title} {\enquote {\bibinfo {title} {Quantum
  chaos, pole-skipping and hydrodynamics in a holographic system with chiral
  anomaly},}\ }\href@noop {} {\bibfield  {journal} {\bibinfo  {journal} {JHEP}\
  }\textbf {\bibinfo {volume} {03}},\ \bibinfo {pages} {050} (\bibinfo {year}
  {2020})}\BibitemShut {NoStop}%
\bibitem [{\citenamefont {Abbasi}(2022)}]{Abbasi_2022}%
  \BibitemOpen
  \bibfield  {author} {\bibinfo {author} {\bibfnamefont {N.}~\bibnamefont
  {Abbasi}},\ }\bibfield  {title} {\enquote {\bibinfo {title} {Long-time tails
  in the syk chain from the effective field theory with a large number of
  derivatives},}\ }\href@noop {} {\bibfield  {journal} {\bibinfo  {journal}
  {JHEP}\ }\textbf {\bibinfo {volume} {04}},\ \bibinfo {pages} {181} (\bibinfo
  {year} {2022})}\BibitemShut {NoStop}%
\bibitem [{\citenamefont {Chen-Lin}\ \emph {et~al.}(2019)\citenamefont
  {Chen-Lin}, \citenamefont {Delacr{\'e}taz},\ and\ \citenamefont
  {Hartnoll}}]{Lin_Delacretaz_Hartnoll_2019}%
  \BibitemOpen
  \bibfield  {author} {\bibinfo {author} {\bibfnamefont {X.}~\bibnamefont
  {Chen-Lin}}, \bibinfo {author} {\bibfnamefont {L.~V.}\ \bibnamefont
  {Delacr{\'e}taz}}, \ and\ \bibinfo {author} {\bibfnamefont {S.~A.}\
  \bibnamefont {Hartnoll}},\ }\bibfield  {title} {\enquote {\bibinfo {title}
  {Theory of diffusive fluctuations},}\ }\href@noop {} {\bibfield  {journal}
  {\bibinfo  {journal} {Phys. Rev. Lett..}\ }\textbf {\bibinfo {volume}
  {122}},\ \bibinfo {pages} {091602} (\bibinfo {year} {2019})}\BibitemShut
  {NoStop}%
\bibitem [{\citenamefont {Weinstock}(1965)}]{Weinstock_1965}%
  \BibitemOpen
  \bibfield  {author} {\bibinfo {author} {\bibfnamefont {J.}~\bibnamefont
  {Weinstock}},\ }\bibfield  {title} {\enquote {\bibinfo {title}
  {Nonanalyticity of transport coefficients and the complete density expansion
  of momentum correlation functions},}\ }\href@noop {} {\bibfield  {journal}
  {\bibinfo  {journal} {Phys. Rev. A}\ }\textbf {\bibinfo {volume} {140}},\
  \bibinfo {pages} {460} (\bibinfo {year} {1965})}\BibitemShut {NoStop}%
\bibitem [{\citenamefont {Dorfman}\ and\ \citenamefont
  {Cohen}(1965)}]{Dorfman_Cohen_1965}%
  \BibitemOpen
  \bibfield  {author} {\bibinfo {author} {\bibfnamefont {J.~R.}\ \bibnamefont
  {Dorfman}}\ and\ \bibinfo {author} {\bibfnamefont {E.~G.~D.}\ \bibnamefont
  {Cohen}},\ }\bibfield  {title} {\enquote {\bibinfo {title} {On the density
  expansion of the pair distribution function for a dense gas not in
  equilibrium},}\ }\href@noop {} {\bibfield  {journal} {\bibinfo  {journal}
  {Phys. Lett.}\ }\textbf {\bibinfo {volume} {16}},\ \bibinfo {pages} {124}
  (\bibinfo {year} {1965})}\BibitemShut {NoStop}%
\bibitem [{\citenamefont {Dorfman}\ and\ \citenamefont
  {Cohen}(1967)}]{Dorfman_Cohen_1967}%
  \BibitemOpen
  \bibfield  {author} {\bibinfo {author} {\bibfnamefont {J.~R.}\ \bibnamefont
  {Dorfman}}\ and\ \bibinfo {author} {\bibfnamefont {E.~G.~D.}\ \bibnamefont
  {Cohen}},\ }\bibfield  {title} {\enquote {\bibinfo {title} {Difficulties in
  the kinetic theory of dense gases},}\ }\href@noop {} {\bibfield  {journal}
  {\bibinfo  {journal} {J. Math. Phys.}\ }\textbf {\bibinfo {volume} {8}},\
  \bibinfo {pages} {282} (\bibinfo {year} {1967})}\BibitemShut {NoStop}%
\bibitem [{\citenamefont {Sengers}(1965)}]{Sengers_1965}%
  \BibitemOpen
  \bibfield  {author} {\bibinfo {author} {\bibfnamefont {J.~V.}\ \bibnamefont
  {Sengers}},\ }\bibfield  {title} {\enquote {\bibinfo {title} {Density
  expansion of the viscosity of a moderately dense gas},}\ }\href@noop {}
  {\bibfield  {journal} {\bibinfo  {journal} {Phys. Rev. Lett.}\ }\textbf
  {\bibinfo {volume} {15}},\ \bibinfo {pages} {515} (\bibinfo {year}
  {1965})}\BibitemShut {NoStop}%
\bibitem [{\citenamefont {Kawasaki}\ and\ \citenamefont
  {Oppenheim}(1965)}]{Kawasaki_Oppenheim_1965}%
  \BibitemOpen
  \bibfield  {author} {\bibinfo {author} {\bibfnamefont {K.}~\bibnamefont
  {Kawasaki}}\ and\ \bibinfo {author} {\bibfnamefont {I.}~\bibnamefont
  {Oppenheim}},\ }\bibfield  {title} {\enquote {\bibinfo {title} {Logarithmic
  terms in the density expansion of transport coefficients},}\ }\href@noop {}
  {\bibfield  {journal} {\bibinfo  {journal} {Phys. Rev.}\ }\textbf {\bibinfo
  {volume} {139A}},\ \bibinfo {pages} {1763} (\bibinfo {year}
  {1965})}\BibitemShut {NoStop}%
\bibitem [{\citenamefont {Anderson}(1958)}]{Anderson_1958}%
  \BibitemOpen
  \bibfield  {author} {\bibinfo {author} {\bibfnamefont {P.~W.}\ \bibnamefont
  {Anderson}},\ }\bibfield  {title} {\enquote {\bibinfo {title} {Absence of
  diffusion in certain random lattices},}\ }\href@noop {} {\bibfield  {journal}
  {\bibinfo  {journal} {Phys. Rev.}\ }\textbf {\bibinfo {volume} {109}},\
  \bibinfo {pages} {1492} (\bibinfo {year} {1958})}\BibitemShut {NoStop}%
\bibitem [{\citenamefont {Kirkpatrick}\ and\ \citenamefont
  {Dorfman}(1983)}]{Kirkpatrick_Dorfman_1983}%
  \BibitemOpen
  \bibfield  {author} {\bibinfo {author} {\bibfnamefont {T.~R.}\ \bibnamefont
  {Kirkpatrick}}\ and\ \bibinfo {author} {\bibfnamefont {J.~R.}\ \bibnamefont
  {Dorfman}},\ }\bibfield  {title} {\enquote {\bibinfo {title} {Divergences and
  long-time tails in two- and three-dimensional quantum lorentz gases},}\
  }\href@noop {} {\bibfield  {journal} {\bibinfo  {journal} {Phys. Rev. A}\
  }\textbf {\bibinfo {volume} {28}},\ \bibinfo {pages} {1022} (\bibinfo {year}
  {1983})}\BibitemShut {NoStop}%
\bibitem [{\citenamefont {Wysoki{\'n}ski}\ \emph {et~al.}(1994)\citenamefont
  {Wysoki{\'n}ski}, \citenamefont {Park}, \citenamefont {Belitz},\ and\
  \citenamefont {Kirkpatrick}}]{Wysokinski_et_al_1994}%
  \BibitemOpen
  \bibfield  {author} {\bibinfo {author} {\bibfnamefont {K.~I.}\ \bibnamefont
  {Wysoki{\'n}ski}}, \bibinfo {author} {\bibfnamefont {W.}~\bibnamefont
  {Park}}, \bibinfo {author} {\bibfnamefont {D.}~\bibnamefont {Belitz}}, \ and\
  \bibinfo {author} {\bibfnamefont {T.~R.}\ \bibnamefont {Kirkpatrick}},\
  }\bibfield  {title} {\enquote {\bibinfo {title} {Density expansion for the
  mobility of electrons in helium gas},}\ }\href@noop {} {\bibfield  {journal}
  {\bibinfo  {journal} {Phys. Rev. Lett.}\ }\textbf {\bibinfo {volume} {73}},\
  \bibinfo {pages} {2571} (\bibinfo {year} {1994})}\BibitemShut {NoStop}%
\bibitem [{\citenamefont {Wysoki{\'n}ski}\ \emph {et~al.}(1995)\citenamefont
  {Wysoki{\'n}ski}, \citenamefont {Park}, \citenamefont {Belitz},\ and\
  \citenamefont {Kirkpatrick}}]{Wysokinski_et_al_1995}%
  \BibitemOpen
  \bibfield  {author} {\bibinfo {author} {\bibfnamefont {K.~I.}\ \bibnamefont
  {Wysoki{\'n}ski}}, \bibinfo {author} {\bibfnamefont {W.}~\bibnamefont
  {Park}}, \bibinfo {author} {\bibfnamefont {D.}~\bibnamefont {Belitz}}, \ and\
  \bibinfo {author} {\bibfnamefont {T.~R.}\ \bibnamefont {Kirkpatrick}},\
  }\bibfield  {title} {\enquote {\bibinfo {title} {Density expansion for the
  mobility in a quantum lorentz model},}\ }\href@noop {} {\bibfield  {journal}
  {\bibinfo  {journal} {Phys. Rev. E}\ }\textbf {\bibinfo {volume} {52}},\
  \bibinfo {pages} {612} (\bibinfo {year} {1995})}\BibitemShut {NoStop}%
\bibitem [{\citenamefont {Dorfman}\ \emph {et~al.}(2021)\citenamefont
  {Dorfman}, \citenamefont {van Beijeren},\ and\ \citenamefont
  {Kirkpatrick}}]{Dorfman_vanBeijeren_Kirkpatrick_2021}%
  \BibitemOpen
  \bibfield  {author} {\bibinfo {author} {\bibfnamefont {J.~R.}\ \bibnamefont
  {Dorfman}}, \bibinfo {author} {\bibfnamefont {H.}~\bibnamefont {van
  Beijeren}}, \ and\ \bibinfo {author} {\bibfnamefont {T.~R.}\ \bibnamefont
  {Kirkpatrick}},\ }\href@noop {} {\emph {\bibinfo {title} {Contemporary
  Kinetic Theory of Matter}}}\ (\bibinfo  {publisher} {Cambridge University
  Press, Cambridge, UK},\ \bibinfo {year} {2021})\BibitemShut {NoStop}%
\bibitem [{\citenamefont {Pomeau}(1972)}]{Pomeau_1972}%
  \BibitemOpen
  \bibfield  {author} {\bibinfo {author} {\bibfnamefont {Y.}~\bibnamefont
  {Pomeau}},\ }\bibfield  {title} {\enquote {\bibinfo {title} {Low-frequency
  behavior of transport coefficients in fluids},}\ }\href@noop {} {\bibfield
  {journal} {\bibinfo  {journal} {Phys. Rev. A}\ }\textbf {\bibinfo {volume}
  {5}},\ \bibinfo {pages} {2569} (\bibinfo {year} {1972})}\BibitemShut
  {NoStop}%
\bibitem [{\citenamefont {Forster}\ \emph {et~al.}(1977)\citenamefont
  {Forster}, \citenamefont {Nelson},\ and\ \citenamefont
  {Stephen}}]{Forster_Nelson_Stephen_1977}%
  \BibitemOpen
  \bibfield  {author} {\bibinfo {author} {\bibfnamefont {D.}~\bibnamefont
  {Forster}}, \bibinfo {author} {\bibfnamefont {D.~R.}\ \bibnamefont {Nelson}},
  \ and\ \bibinfo {author} {\bibfnamefont {M.~J.}\ \bibnamefont {Stephen}},\
  }\bibfield  {title} {\enquote {\bibinfo {title} {Large-distance and long-time
  properties of a randomly stirred fluid},}\ }\href@noop {} {\bibfield
  {journal} {\bibinfo  {journal} {Phys. Rev. A}\ }\textbf {\bibinfo {volume}
  {16}},\ \bibinfo {pages} {732} (\bibinfo {year} {1977})}\BibitemShut
  {NoStop}%
\bibitem [{\citenamefont {Bergmann}(1984)}]{Bergmann_1984}%
  \BibitemOpen
  \bibfield  {author} {\bibinfo {author} {\bibfnamefont {G.}~\bibnamefont
  {Bergmann}},\ }\bibfield  {title} {\enquote {\bibinfo {title} {Weak
  localization in thin films: a time-of-flight experiment with conduction
  electrons},}\ }\href@noop {} {\bibfield  {journal} {\bibinfo  {journal}
  {Phys. Rep.}\ }\textbf {\bibinfo {volume} {107}},\ \bibinfo {pages} {1}
  (\bibinfo {year} {1984})}\BibitemShut {NoStop}%
\bibitem [{\citenamefont {Abrahams}\ \emph {et~al.}(1979)\citenamefont
  {Abrahams}, \citenamefont {Anderson}, \citenamefont {Licciardello},\ and\
  \citenamefont {Ramakrishnan}}]{Abrahams_et_al_1979}%
  \BibitemOpen
  \bibfield  {author} {\bibinfo {author} {\bibfnamefont {E.}~\bibnamefont
  {Abrahams}}, \bibinfo {author} {\bibfnamefont {P.~W.}\ \bibnamefont
  {Anderson}}, \bibinfo {author} {\bibfnamefont {D.~C.}\ \bibnamefont
  {Licciardello}}, \ and\ \bibinfo {author} {\bibfnamefont {T.~V.}\
  \bibnamefont {Ramakrishnan}},\ }\bibfield  {title} {\enquote {\bibinfo
  {title} {Scaling theory of localization: Absence of quantum diffusion in two
  dimensions},}\ }\href@noop {} {\bibfield  {journal} {\bibinfo  {journal}
  {Phys. Rev. Lett.}\ }\textbf {\bibinfo {volume} {42}},\ \bibinfo {pages}
  {673} (\bibinfo {year} {1979})}\BibitemShut {NoStop}%
\bibitem [{\citenamefont {Wegner}(1979)}]{Wegner_1979}%
  \BibitemOpen
  \bibfield  {author} {\bibinfo {author} {\bibfnamefont {F.}~\bibnamefont
  {Wegner}},\ }\bibfield  {title} {\enquote {\bibinfo {title} {The mobility
  edge problem: Continuous symmetry and a conjecture},}\ }\href@noop {}
  {\bibfield  {journal} {\bibinfo  {journal} {Z. Phys. B}\ }\textbf {\bibinfo
  {volume} {35}},\ \bibinfo {pages} {207} (\bibinfo {year} {1979})}\BibitemShut
  {NoStop}%
\bibitem [{\citenamefont {Lee}\ and\ \citenamefont
  {Ramakrishnan}(1985)}]{Lee_Ramakrishnan_1985}%
  \BibitemOpen
  \bibfield  {author} {\bibinfo {author} {\bibfnamefont {P.~A.}\ \bibnamefont
  {Lee}}\ and\ \bibinfo {author} {\bibfnamefont {T.~V.}\ \bibnamefont
  {Ramakrishnan}},\ }\bibfield  {title} {\enquote {\bibinfo {title} {Disordered
  electronic systems},}\ }\href@noop {} {\bibfield  {journal} {\bibinfo
  {journal} {Rev. Mod. Phys.}\ }\textbf {\bibinfo {volume} {57}},\ \bibinfo
  {pages} {287} (\bibinfo {year} {1985})}\BibitemShut {NoStop}%
\bibitem [{\citenamefont {Evers}\ and\ \citenamefont
  {Mirlin}(2008)}]{Evers_Mirlin_2008}%
  \BibitemOpen
  \bibfield  {author} {\bibinfo {author} {\bibfnamefont {F.}~\bibnamefont
  {Evers}}\ and\ \bibinfo {author} {\bibfnamefont {A.~D.}\ \bibnamefont
  {Mirlin}},\ }\bibfield  {title} {\enquote {\bibinfo {title} {Anderson
  transitions},}\ }\href@noop {} {\bibfield  {journal} {\bibinfo  {journal}
  {Rev. Mod. Phys.}\ }\textbf {\bibinfo {volume} {80}},\ \bibinfo {pages}
  {1355} (\bibinfo {year} {2008})}\BibitemShut {NoStop}%
\bibitem [{\citenamefont {Ernst}\ \emph {et~al.}(1971)\citenamefont {Ernst},
  \citenamefont {Hauge},\ and\ \citenamefont {van
  Leeuwen}}]{Ernst_Hauge_van_Leeuwen_1971}%
  \BibitemOpen
  \bibfield  {author} {\bibinfo {author} {\bibfnamefont {M.~H.}\ \bibnamefont
  {Ernst}}, \bibinfo {author} {\bibfnamefont {E.~H.}\ \bibnamefont {Hauge}}, \
  and\ \bibinfo {author} {\bibfnamefont {J.~M.~J.}\ \bibnamefont {van
  Leeuwen}},\ }\bibfield  {title} {\enquote {\bibinfo {title} {Asymptotic time
  behavior of correlation functions. {I}. kinetic terms},}\ }\href@noop {}
  {\bibfield  {journal} {\bibinfo  {journal} {Phys. Rev. A}\ }\textbf {\bibinfo
  {volume} {4}},\ \bibinfo {pages} {2055} (\bibinfo {year} {1971})}\BibitemShut
  {NoStop}%
\bibitem [{\citenamefont {G{\"o}tze}(1979)}]{Goetze_1979}%
  \BibitemOpen
  \bibfield  {author} {\bibinfo {author} {\bibfnamefont {W.}~\bibnamefont
  {G{\"o}tze}},\ }\bibfield  {title} {\enquote {\bibinfo {title} {A theory for
  the conductivity of a fermion gas moving in a strong three-dimensional random
  potential},}\ }\href@noop {} {\bibfield  {journal} {\bibinfo  {journal} {J.
  Phys. C: Solid State Phys.}\ }\textbf {\bibinfo {volume} {12}},\ \bibinfo
  {pages} {1279} (\bibinfo {year} {1979})}\BibitemShut {NoStop}%
\bibitem [{\citenamefont {G{\"o}tze}\ \emph {et~al.}(1981)\citenamefont
  {G{\"o}tze}, \citenamefont {Leutheusser},\ and\ \citenamefont
  {Yip}}]{Goetze_Leutheusser_Yip_1981}%
  \BibitemOpen
  \bibfield  {author} {\bibinfo {author} {\bibfnamefont {W.}~\bibnamefont
  {G{\"o}tze}}, \bibinfo {author} {\bibfnamefont {E.}~\bibnamefont
  {Leutheusser}}, \ and\ \bibinfo {author} {\bibfnamefont {S.}~\bibnamefont
  {Yip}},\ }\bibfield  {title} {\enquote {\bibinfo {title} {Dynamical theory of
  diffusion and localization in a random, static field},}\ }\href@noop {}
  {\bibfield  {journal} {\bibinfo  {journal} {Phys. Rev. A}\ }\textbf {\bibinfo
  {volume} {23}},\ \bibinfo {pages} {2634} (\bibinfo {year}
  {1981})}\BibitemShut {NoStop}%
\bibitem [{\citenamefont {G{\"o}tze}(1981)}]{Goetze_1981}%
  \BibitemOpen
  \bibfield  {author} {\bibinfo {author} {\bibfnamefont {W.}~\bibnamefont
  {G{\"o}tze}},\ }\bibfield  {title} {\enquote {\bibinfo {title} {The mobility
  of a quantum particle in a three-dimensional random potential},}\ }\href@noop
  {} {\bibfield  {journal} {\bibinfo  {journal} {Phil. Mag. B}\ }\textbf
  {\bibinfo {volume} {43}},\ \bibinfo {pages} {219} (\bibinfo {year}
  {1981})}\BibitemShut {NoStop}%
\bibitem [{\citenamefont {Gorkov}\ \emph {et~al.}(1979)\citenamefont {Gorkov},
  \citenamefont {Larkin},\ and\ \citenamefont
  {Khmelnitskii}}]{Gorkov_Larkin_Khmelnitskii_1979}%
  \BibitemOpen
  \bibfield  {author} {\bibinfo {author} {\bibfnamefont {L.~P.}\ \bibnamefont
  {Gorkov}}, \bibinfo {author} {\bibfnamefont {A.}~\bibnamefont {Larkin}}, \
  and\ \bibinfo {author} {\bibfnamefont {D.~E.}\ \bibnamefont {Khmelnitskii}},\
  }\bibfield  {title} {\enquote {\bibinfo {title} {Particle conductivity in a
  two-dimensional random potential},}\ }\href@noop {} {\bibfield  {journal}
  {\bibinfo  {journal} {Pis'ma Zh. Eksp. Teor. Fiz.}\ }\textbf {\bibinfo
  {volume} {30}},\ \bibinfo {pages} {248} (\bibinfo {year} {1979})},\ \bibinfo
  {note} {[JETP Lett. {\bf 30}, 228 (1979)}\BibitemShut {NoStop}%
\bibitem [{\citenamefont {Vollhardt}\ and\ \citenamefont
  {W{\"o}lfle}(1980)}]{Vollhardt_Woelfle_1980}%
  \BibitemOpen
  \bibfield  {author} {\bibinfo {author} {\bibfnamefont {D.}~\bibnamefont
  {Vollhardt}}\ and\ \bibinfo {author} {\bibfnamefont {P.}~\bibnamefont
  {W{\"o}lfle}},\ }\bibfield  {title} {\enquote {\bibinfo {title}
  {Diagrammatic, self-consistent treatment of the anderson localization problem
  in $d\leq 2$ dimensions},}\ }\href@noop {} {\bibfield  {journal} {\bibinfo
  {journal} {Phys. Rev. B}\ }\textbf {\bibinfo {volume} {22}},\ \bibinfo
  {pages} {4666} (\bibinfo {year} {1980})}\BibitemShut {NoStop}%
\bibitem [{\citenamefont {Sch{\"a}fer}\ and\ \citenamefont
  {Wegner}(1980)}]{Schaefer_Wegner_1980}%
  \BibitemOpen
  \bibfield  {author} {\bibinfo {author} {\bibfnamefont {L.}~\bibnamefont
  {Sch{\"a}fer}}\ and\ \bibinfo {author} {\bibfnamefont {F.}~\bibnamefont
  {Wegner}},\ }\bibfield  {title} {\enquote {\bibinfo {title} {Disordered
  system with n orbitals per site: Lagrange formulation, hyperbolic symmetry,
  and goldstone modes},}\ }\href@noop {} {\bibfield  {journal} {\bibinfo
  {journal} {Z. Phys. B}\ }\textbf {\bibinfo {volume} {38}},\ \bibinfo {pages}
  {113} (\bibinfo {year} {1980})}\BibitemShut {NoStop}%
\bibitem [{\citenamefont {Lighthill}(1958)}]{Lighthill_1958}%
  \BibitemOpen
  \bibfield  {author} {\bibinfo {author} {\bibfnamefont {M.~J.}\ \bibnamefont
  {Lighthill}},\ }\href@noop {} {\emph {\bibinfo {title} {Introduction to
  Fourier analysis and generalised functions}}}\ (\bibinfo  {publisher}
  {Cambridge University Press, Cambridge},\ \bibinfo {year} {1958})\BibitemShut
  {NoStop}%
\bibitem [{\citenamefont {Feller}(1970)}]{Feller_1970}%
  \BibitemOpen
  \bibfield  {author} {\bibinfo {author} {\bibfnamefont {W.}~\bibnamefont
  {Feller}},\ }\href@noop {} {\emph {\bibinfo {title} {An Introduction to
  Probability Theory and Its Applications}}},\ Vol.~\bibinfo {volume} {II}\
  (\bibinfo  {publisher} {Wiley, New York},\ \bibinfo {year}
  {1970})\BibitemShut {NoStop}%
\bibitem [{QM_()}]{QM_footnote}%
  \BibitemOpen
  \href@noop {} {}\bibinfo {note} {At this point, this is the only way in which
  quantum mechanics enters explicitly. In Sec.~\ref{sec:IV} we will see that
  for finding the leading quantum mechanical LTT one also needs to consider
  quantum mechanical manifestations of the diffusion propagator ${\cal
  D}$.}\BibitemShut {Stop}%
\bibitem [{typ()}]{typo_footnote}%
  \BibitemOpen
  \href@noop {} {}\bibinfo {note} {Equation (7) in
  Ref.~\onlinecite{Ernst_Weyland_1971} contains a typographical error: The
  prefactor of the integral in Eq.~(7) should read $v^2/n\,d^2$.}\BibitemShut
  {Stop}%
\bibitem [{\citenamefont {Forster}(1975)}]{Forster_1975}%
  \BibitemOpen
  \bibfield  {author} {\bibinfo {author} {\bibfnamefont {D.}~\bibnamefont
  {Forster}},\ }\href@noop {} {\emph {\bibinfo {title} {Hydrodynamic
  Fluctuations, Broken Symmetry, and Correlation Functions}}}\ (\bibinfo
  {publisher} {Benjamin, Reading, MA},\ \bibinfo {year} {1975})\BibitemShut
  {NoStop}%
\bibitem [{not()}]{notation_footnote}%
  \BibitemOpen
  \href@noop {} {}\bibinfo {note} {In contrast to Appendix~\ref{app:A} we
  denote the time-correlation function and its temporal Fourier and Laplace
  transforms by the same symbol, and distinguish between them by their
  respective arguments.}\BibitemShut {Stop}%
\bibitem [{uni()}]{uniqueness_footnote}%
  \BibitemOpen
  \href@noop {} {}\bibinfo {note} {{\it A priori} it is not obvious that this
  is the only possible QM generalization of Eq.~(\ref{eq:3.15}), as one could
  also replace the van Hove function with the anticommutator, or symmetrized,
  density correlation function. However, the same diffusive propagator ${\cal
  D}$ governs the decay of a density perturbation, see Appendix~\ref{app:B.3},
  and the relevant QM relaxation function is the Kubo function, see
  Ref.~\onlinecite{Kubo_1957}. This makes the QM identification of ${\cal D}$
  in terms of $\Phi_{nn}$ unique.}\BibitemShut {Stop}%
\bibitem [{WL_()}]{WL_prefactor_footnote}%
  \BibitemOpen
  \href@noop {} {}\bibinfo {note} {This can be seen most easily by writing
  Eq.~(42) in Ref.~\onlinecite{Vollhardt_Woelfle_1980} in terms of the
  diffusion coefficient, and comparing the result with
  Eq.~(\ref{eq:4.12}).}\BibitemShut {Stop}%
\bibitem [{\citenamefont {Prelov\v{s}ek}(1981)}]{Prelovsek_1981}%
  \BibitemOpen
  \bibfield  {author} {\bibinfo {author} {\bibfnamefont {P.}~\bibnamefont
  {Prelov\v{s}ek}},\ }\bibfield  {title} {\enquote {\bibinfo {title}
  {Conductor-insulator transition in the anderson model of a disordered
  solid},}\ }\href@noop {} {\bibfield  {journal} {\bibinfo  {journal} {Phys.
  Rev. B}\ }\textbf {\bibinfo {volume} {23}},\ \bibinfo {pages} {1304}
  (\bibinfo {year} {1981})}\BibitemShut {NoStop}%
\bibitem [{\citenamefont {Belitz}\ \emph {et~al.}(1981)\citenamefont {Belitz},
  \citenamefont {Gold},\ and\ \citenamefont
  {G{\"o}tze}}]{Belitz_Gold_Goetze_1981}%
  \BibitemOpen
  \bibfield  {author} {\bibinfo {author} {\bibfnamefont {D.}~\bibnamefont
  {Belitz}}, \bibinfo {author} {\bibfnamefont {A.}~\bibnamefont {Gold}}, \ and\
  \bibinfo {author} {\bibfnamefont {W.}~\bibnamefont {G{\"o}tze}},\ }\bibfield
  {title} {\enquote {\bibinfo {title} {Self-consistent current relaxation
  theory for the electron localization problem},}\ }\href@noop {} {\bibfield
  {journal} {\bibinfo  {journal} {Z. Phys. B}\ }\textbf {\bibinfo {volume}
  {44}},\ \bibinfo {pages} {273} (\bibinfo {year} {1981})}\BibitemShut
  {NoStop}%
\bibitem [{\citenamefont {Wegner}(1989)}]{Wegner_1989}%
  \BibitemOpen
  \bibfield  {author} {\bibinfo {author} {\bibfnamefont {F.}~\bibnamefont
  {Wegner}},\ }\bibfield  {title} {\enquote {\bibinfo {title} {Four-loop-order
  $\beta$-function of nonlinear $\sigma$-models in symmetric spaces},}\
  }\href@noop {} {\bibfield  {journal} {\bibinfo  {journal} {Nucl. Phys. B}\
  }\textbf {\bibinfo {volume} {316}},\ \bibinfo {pages} {663} (\bibinfo {year}
  {1989})}\BibitemShut {NoStop}%
\bibitem [{\citenamefont {Belitz}\ and\ \citenamefont
  {Yang}(1993)}]{Belitz_Yang_1993}%
  \BibitemOpen
  \bibfield  {author} {\bibinfo {author} {\bibfnamefont {D.}~\bibnamefont
  {Belitz}}\ and\ \bibinfo {author} {\bibfnamefont {S.~Q.}\ \bibnamefont
  {Yang}},\ }\bibfield  {title} {\enquote {\bibinfo {title} {Scaling functions
  and equations of state for nonlinear sigma models},}\ }\href@noop {}
  {\bibfield  {journal} {\bibinfo  {journal} {Nucl. Phys. B}\ }\textbf
  {\bibinfo {volume} {401}},\ \bibinfo {pages} {548} (\bibinfo {year}
  {1993})}\BibitemShut {NoStop}%
\bibitem [{\citenamefont {Alder}\ and\ \citenamefont
  {Alley}(1978)}]{Alder_Alley_1978}%
  \BibitemOpen
  \bibfield  {author} {\bibinfo {author} {\bibfnamefont {B.~J.}\ \bibnamefont
  {Alder}}\ and\ \bibinfo {author} {\bibfnamefont {W.~E.}\ \bibnamefont
  {Alley}},\ }\bibfield  {title} {\enquote {\bibinfo {title} {Long-time
  correlation effects on displacement distributions},}\ }\href@noop {}
  {\bibfield  {journal} {\bibinfo  {journal} {J. Stat. Phys.}\ }\textbf
  {\bibinfo {volume} {19}},\ \bibinfo {pages} {341} (\bibinfo {year}
  {1978})}\BibitemShut {NoStop}%
\bibitem [{\citenamefont {H{\"o}fling}\ \emph {et~al.}(2006)\citenamefont
  {H{\"o}fling}, \citenamefont {Franosch},\ and\ \citenamefont
  {Frey}}]{Hoefling_Franosch_Frey_2006}%
  \BibitemOpen
  \bibfield  {author} {\bibinfo {author} {\bibfnamefont {F.}~\bibnamefont
  {H{\"o}fling}}, \bibinfo {author} {\bibfnamefont {T.}~\bibnamefont
  {Franosch}}, \ and\ \bibinfo {author} {\bibfnamefont {E.}~\bibnamefont
  {Frey}},\ }\bibfield  {title} {\enquote {\bibinfo {title} {Localization
  transition of the three-dimensional lorentz model and continuum
  percolation},}\ }\href@noop {} {\bibfield  {journal} {\bibinfo  {journal}
  {Phys. Rev. Lett.}\ }\textbf {\bibinfo {volume} {96}},\ \bibinfo {pages}
  {165901} (\bibinfo {year} {2006})}\BibitemShut {NoStop}%
\bibitem [{\citenamefont {Ernst}\ and\ \citenamefont
  {Dorfman}(1975)}]{Ernst_Dorfman_1975}%
  \BibitemOpen
  \bibfield  {author} {\bibinfo {author} {\bibfnamefont {M.~H.}\ \bibnamefont
  {Ernst}}\ and\ \bibinfo {author} {\bibfnamefont {J.~R.}\ \bibnamefont
  {Dorfman}},\ }\bibfield  {title} {\enquote {\bibinfo {title} {Nonanalytic
  dispersion relations for classical fluids},}\ }\href@noop {} {\bibfield
  {journal} {\bibinfo  {journal} {J. Stat. Phys.}\ }\textbf {\bibinfo {volume}
  {12}},\ \bibinfo {pages} {311} (\bibinfo {year} {1975})}\BibitemShut
  {NoStop}%
\bibitem [{\citenamefont {Altshuler}\ and\ \citenamefont
  {Aronov}(1985)}]{Altshuler_Aronov_1985}%
  \BibitemOpen
  \bibfield  {author} {\bibinfo {author} {\bibfnamefont {B.~L.}\ \bibnamefont
  {Altshuler}}\ and\ \bibinfo {author} {\bibfnamefont {A.~G.}\ \bibnamefont
  {Aronov}},\ }\bibfield  {title} {\enquote {\bibinfo {title}
  {Electron-electron interaction in disordered conductors},}\ }in\ \href@noop
  {} {\emph {\bibinfo {booktitle} {Electron-Electron Interactions in Disordered
  Systems}}},\ \bibinfo {editor} {edited by\ \bibinfo {editor} {\bibfnamefont
  {A.~L.}\ \bibnamefont {Efros}}\ and\ \bibinfo {editor} {\bibfnamefont
  {M.}~\bibnamefont {Pollak}}}\ (\bibinfo  {publisher} {North-Holland,
  Amsterdam},\ \bibinfo {year} {1985})\ p.~\bibinfo {pages} {1}\BibitemShut
  {NoStop}%
\bibitem [{\citenamefont {Belitz}\ and\ \citenamefont
  {Kirkpatrick}(1994)}]{Belitz_Kirkpatrick_1994}%
  \BibitemOpen
  \bibfield  {author} {\bibinfo {author} {\bibfnamefont {D.}~\bibnamefont
  {Belitz}}\ and\ \bibinfo {author} {\bibfnamefont {T.~R.}\ \bibnamefont
  {Kirkpatrick}},\ }\bibfield  {title} {\enquote {\bibinfo {title} {The
  {A}nderson-{M}ott transition},}\ }\href@noop {} {\bibfield  {journal}
  {\bibinfo  {journal} {Rev. Mod. Phys.}\ }\textbf {\bibinfo {volume} {66}},\
  \bibinfo {pages} {261} (\bibinfo {year} {1994})}\BibitemShut {NoStop}%
\bibitem [{\citenamefont {Altshuler}\ \emph {et~al.}(1980)\citenamefont
  {Altshuler}, \citenamefont {Aronov},\ and\ \citenamefont
  {Lee}}]{Altshuler_Aronov_Lee_1980}%
  \BibitemOpen
  \bibfield  {author} {\bibinfo {author} {\bibfnamefont {B.~L.}\ \bibnamefont
  {Altshuler}}, \bibinfo {author} {\bibfnamefont {A.~G.}\ \bibnamefont
  {Aronov}}, \ and\ \bibinfo {author} {\bibfnamefont {P.~A.}\ \bibnamefont
  {Lee}},\ }\bibfield  {title} {\enquote {\bibinfo {title} {Interaction effect
  in disordered fermi systems in two dimensions},}\ }\href@noop {} {\bibfield
  {journal} {\bibinfo  {journal} {Phys. Rev. Lett.}\ }\textbf {\bibinfo
  {volume} {44}},\ \bibinfo {pages} {1288} (\bibinfo {year}
  {1980})}\BibitemShut {NoStop}%
\bibitem [{cla()}]{classical_Stat_Mech_footnote}%
  \BibitemOpen
  \href@noop {} {}\bibinfo {note} {In Eqs.~(\ref{eqs:B.1}) and (\ref{eq:B.2})
  we use classical Statistical Mechanics, but the result is valid for quantum
  Lorentz models as well.}\BibitemShut {Stop}%
\bibitem [{mod()}]{mode_coupling_footnote}%
  \BibitemOpen
  \href@noop {} {}\bibinfo {note} {There is no universal agreement about the
  definition of the term `mode coupling theory'. We use it for theories that
  are in the spirit of Refs.~\onlinecite{Kadanoff_Swift_1968} and
  \onlinecite{Kawasaki_1970}.}\BibitemShut {Stop}%
\bibitem [{\citenamefont {Kadanoff}\ and\ \citenamefont
  {Swift}(1968)}]{Kadanoff_Swift_1968}%
  \BibitemOpen
  \bibfield  {author} {\bibinfo {author} {\bibfnamefont {L.~P.}\ \bibnamefont
  {Kadanoff}}\ and\ \bibinfo {author} {\bibfnamefont {J.}~\bibnamefont
  {Swift}},\ }\bibfield  {title} {\enquote {\bibinfo {title} {Transport
  coefficients near the liquid-gas critical point},}\ }\href@noop {} {\bibfield
   {journal} {\bibinfo  {journal} {Phys. Rev.}\ }\textbf {\bibinfo {volume}
  {166}},\ \bibinfo {pages} {89} (\bibinfo {year} {1968})}\BibitemShut
  {NoStop}%
\bibitem [{\citenamefont {Kawasaki}(1970)}]{Kawasaki_1970}%
  \BibitemOpen
  \bibfield  {author} {\bibinfo {author} {\bibfnamefont {K.}~\bibnamefont
  {Kawasaki}},\ }\bibfield  {title} {\enquote {\bibinfo {title} {Kinetic
  equations and time correlation functions of critical fluctuations},}\
  }\href@noop {} {\bibfield  {journal} {\bibinfo  {journal} {Ann. Phys. (NY)}\
  }\textbf {\bibinfo {volume} {61}},\ \bibinfo {pages} {1} (\bibinfo {year}
  {1970})}\BibitemShut {NoStop}%
\bibitem [{\citenamefont {Zwanzig}(1961)}]{Zwanzig_1961}%
  \BibitemOpen
  \bibfield  {author} {\bibinfo {author} {\bibfnamefont {R.}~\bibnamefont
  {Zwanzig}},\ }in\ \href@noop {} {\emph {\bibinfo {booktitle} {Lectures in
  Theoretical Physics}}},\ Vol.~\bibinfo {volume} {3},\ \bibinfo {editor}
  {edited by\ \bibinfo {editor} {\bibfnamefont {B.~W.~Downs}\ \bibnamefont
  {W.~E.~Brittin}}\ and\ \bibinfo {editor} {\bibfnamefont {J.}~\bibnamefont
  {Downs}}}\ (\bibinfo  {publisher} {Interscience, New York},\ \bibinfo {year}
  {1961})\BibitemShut {NoStop}%
\bibitem [{\citenamefont {Mori}(1965)}]{Mori_1965}%
  \BibitemOpen
  \bibfield  {author} {\bibinfo {author} {\bibfnamefont {H.}~\bibnamefont
  {Mori}},\ }\bibfield  {title} {\enquote {\bibinfo {title} {Transport,
  collective motion, and brownian motion},}\ }\href@noop {} {\bibfield
  {journal} {\bibinfo  {journal} {Progr. Theor. Phys.}\ }\textbf {\bibinfo
  {volume} {33}},\ \bibinfo {pages} {423} (\bibinfo {year} {1965})}\BibitemShut
  {NoStop}%
\bibitem [{\citenamefont {G{\"o}tze}\ and\ \citenamefont
  {W{\"o}lfle}(1972)}]{Goetze_Woefle_1972}%
  \BibitemOpen
  \bibfield  {author} {\bibinfo {author} {\bibfnamefont {W.}~\bibnamefont
  {G{\"o}tze}}\ and\ \bibinfo {author} {\bibfnamefont {P.}~\bibnamefont
  {W{\"o}lfle}},\ }\bibfield  {title} {\enquote {\bibinfo {title} {Homogeneous
  dynamical conductivity of simple metals},}\ }\href@noop {} {\bibfield
  {journal} {\bibinfo  {journal} {Phys. Rev. B}\ }\textbf {\bibinfo {volume}
  {6}},\ \bibinfo {pages} {1226} (\bibinfo {year} {1972})}\BibitemShut
  {NoStop}%
\bibitem [{\citenamefont {Kubo}(1957)}]{Kubo_1957}%
  \BibitemOpen
  \bibfield  {author} {\bibinfo {author} {\bibfnamefont {R.}~\bibnamefont
  {Kubo}},\ }\bibfield  {title} {\enquote {\bibinfo {title}
  {Statistical-mechanical theory of irreversible processes. {I}. general theory
  and simple applications to magnetic and conduction problems},}\ }\href@noop
  {} {\bibfield  {journal} {\bibinfo  {journal} {J. Phys. Soc. Jpn.}\ }\textbf
  {\bibinfo {volume} {12}},\ \bibinfo {pages} {570} (\bibinfo {year}
  {1957})}\BibitemShut {NoStop}%
\end{thebibliography}
%
\end{document}